\documentclass{aa}  
\usepackage{dcolumn}
\makeatletter
 \def\hlinewd#1{%
   \noalign{\ifnum0=`}\fi\hrule \@height #1 \futurelet
    \reserved@a\@xhline}
\makeatother

\newcommand{\kms}{km\,s$^{-1}$}
\begin{document} 

    \title{The peculiar optical-UV X-ray spectra
     of the  X-ray weak quasar PG\,0043+039 
\thanks{Based on observations obtained with the XMM-Newton,
 Hubble Space Telescope (HST), Southern African Large Telescope (SALT),
and Hobby-Eberly Telescope (HET).}
}

  \author{W. Kollatschny \inst{1}, 
           N. Schartel \inst{2},
           M. Zetzl \inst{1}, 
           M. Santos-Lle\'{o} \inst{2},
           P. M. Rodr\'{i}guez-Pascual  \inst{2},
           L. Ballo \inst{3},
           A. Talavera \inst{2}
          }

   \institute{Institut f\"ur Astrophysik, Universit\"at G\"ottingen,
              Friedrich-Hund Platz 1, D-37077 G\"ottingen, Germany\\
              \email{wkollat@astro.physik.uni-goettingen.de} 
            \and
                XMM-Newton Science Operations Centre, ESA, Villafranca del Castillo, Apartado 78, 
              28691 Villanueva de la Ca{\~nada}, Spain 
             \and 
         Osservatorio Astronomico di Brera (INAF), via Brera 28, I-20121 Milano, Italy
}

   \date{Received June 23, 2015; accepted September 28, 2015}
  \authorrunning{Kollatschny et al.}
   \titlerunning{Properties of X-ray weakest quasar PG\,0043+039}

% \abstract{}{}{}{}{} 
% 5 {} token are mandatory
 
  \abstract
  % context heading (optional)
   {The object PG\,0043+039 has been identified as a broad absorption
    line (BAL) quasar  based on its UV spectra.
  However, this optical luminous quasar has not been detected before in deep
   X-ray observations, making it the most extreme X-ray weak quasar known
today.}
  % aims heading (mandatory)
   {This study aims\ to detect PG\,0043+039 in a deep X-ray exposure. The
 question is what causes the extreme X-ray weakness of PG\,0043+039? Does 
 PG\,0043+039 show other spectral or continuum peculiarities?}
  % methods heading (mandatory)
   {We took simultaneous deep X-ray spectra  with XMM-Newton,
    far-ultraviolet (FUV) spectra with the Hubble Space Telescope (HST),
 and optical spectra of PG\,0043+039 with the 
 Hobby-Eberly Telescope (HET) and Southern African Large Telescope (SALT)
 in July, 2013.}  
  % results heading (mandatory)
   {We have  detected PG\,0043+039  in our X-ray exposure taken in 2013.
We presented our first results  in a separate paper
(Kollatschny et al.\citealt{kollatschny15}).
PG\,0043+039 shows an extreme  $\alpha_{ox}$ gradient
($\alpha_{ox}$=$-$2.37). Furthermore, we were able to verify an X-ray flux
of this source 
in a reanalysis of the X-ray data taken in 2005. At that time, it was fainter
by a factor of 3.8~$\pm$0.9 with  $\alpha_{ox}$=$-$2.55.
The X-ray spectrum is compatible with a normal quasar power-law spectrum
($\Gamma$~$=$~1.70$_{-0.45}^{+0.57}$) with moderate intrinsic absorption
(N$_H$~$=$~$5.5_{-3.9}^{+6.9}$~$\times$10$^{21}$~cm$^{-2}$) and reflection. 
The UV/optical flux of PG\,0043+039 has increased by a factor of 1.8
compared to spectra taken in the years 1990-1991. The FUV spectrum
is highly peculiar and dominated by
broad bumps besides Ly$\alpha$. 
 There is no detectable Lyman
edge associated with the BAL absorbing gas seen in the CIV line.
PG\,0043+039 shows a maximum in the overall continuum flux at around 
$\lambda  \approx 2500$ \AA{}  
in contrast to most other AGN where the maximum is found at 
shorter wavelengths.
All the above is compatible with an intrinsically X-ray weak quasar, rather
than an absorbed X-ray emission.
Besides strong FeII multiplets and broad Balmer and HeI lines in the
optical band we only detect 
 a narrow  [\ion{O}{ii}]$\lambda$3727 emission line  and a   
BAL system in the CaH\,$\lambda 3968$, CaK\,$\lambda 3934$ lines
(blueshifted by 4900 kms$^{-1}$) and in the  
\ion{He}{i}\,$\lambda 3889$ line (blueshifted by 5600 kms$^{-1}$).}
{}


   \maketitle
%
%________________________________________________________________

\section{Introduction}

PG\,0043+039 is a bright (m$_v\sim15.5$) and
 luminous quasar (M$_B=-26.11$) at a redshift of
z=0.38512.
A luminosity of $\nu L_{\nu}$~=~2.21 $\times
 10^{44}$~erg~s$^{-1}$  at 3000~\AA\
has been determined before for this quasar
(Baskin \& Laor\citealt{baskin04}). This corresponds to 
an Eddington luminosity log L/$L_{edd}$~=~-0.648 for a black hole mass of
$M = 8.9 \times 10^{9} M_{\odot}$
(Baskin \& Laor\citealt{baskin05}).
A first optical spectrum of PG\,0043+039 taken in Sept. 1990
 has been published by Boroson \& Green\cite{boroson92}.
 PG\,0043+039 has been identified by Bahcall et al.\cite{bahcall93}
and Turnshek et al.\cite{turnshek94,turnshek97}
 as a weak broad absorption line (BAL) quasar
based on CIV BAL detected with the Hubble Space Telescope (HST).
Bechtold et al.\cite{bechtold02} describe the CIV absorber 
in a reanalysis of the HST Faint Object Spectrograph (FOS) spectra
as a very narrow associated absorber.
PG\,0043+039 shows strong Fe II blends in the optical. 
No narrow [OIII], [OII] lines have been detected by Turnshek et
al.\cite{turnshek94} before. The observed continuum in the UV is atypical
in the sense that it is much weaker
relative to the optical continuum compared to normal quasars.
%indicating strong intrinsic reddening by dust.
%On the other hand
There is no evidence for a BAL caused by low-ionization
transitions of, for example, AlII or CII.

A ROSAT nondetection established PG\,0043+039 as an X-ray weak quasar
(Brandt et al.\citealt{brandt00}). 
It was not\ detected in pointed observations
with the ASCA satellite in the year 1996
(Gallagher et al.\citealt{gallagher99}). Furthermore,
PG\,0043+039 is the only quasar in the PG sample 
(Schmidt \& Green,\citealt{schmidt83}),
which was not detected in a dedicated XMM-Newton pointing
(Czerny et al.\citealt{czerny08}).

PG\,0043+039 is the most extreme X-ray weak quasar known to date, but
surprisingly  only shows a  weak BAL system.
The majority of BAL quasars are X-ray weak 
because of a shielding gas (following Murray et al. \citealt{murray95}) or
an intrinsic X-ray weakness that produces more preferable conditions for
wind launching and driving (e.g., Baskin et al. \citealt{baskin13}).
%is usually explained by the absorption of the out-flowing wind
%in combination with the winds velocity shear.
A conclusive interpretation about the X-ray weakness in PG\,0043+039
 was hampered by the absence of simultaneous
measurements, which is mandatory as both the X-ray flux and the BAL 
system are known to be variable. 

There is the possibility that the X-ray quietness of PG\,0043+039
was only a temporal event
or that it was a false conclusion driven by spectra that were not taken  simultaneously in the optical/UV
and X-ray bands. Fluctuations in the accretion process
may have caused a short interruption or huge drop in the X-ray emission as seen
in several low state observations. Examples of the latter are
PG2112+059  (Schartel et al.\citealt{schartel10}) and 1H0707-495
(Fabian et al.\citealt{fabian11}). The source spectroscopic type can
change with time. This was, for example, observed for Fairall\,9 (Sey 1 to
Sey 2; Kollatschny et al.\citealt{kollatschny85}) or NGC\,2617 
(Sey 1.8 to Sey 1; Shappee et al.\citealt{shappee14}). 
The X-ray spectral characteristics can change with time because of
changes in the column density and/or the ionization state of the
X-ray absorbing material. Examples are NGC\,1365 
(Risaliti et al.\citealt{risaliti05}) and NGC\,7583 (Bianchi et al.\citealt{bianchi09}).
BAL systems may develop with time as reported for
WPVS\,007 (Leighly et al.\citealt{leighly09}), where the change in the broad
UV absorption system was reflected in changes of the X-ray brightness
interpreted as due to both X-ray absorption and X-ray weakness.
Here we want to verify that the X-ray weakness
in PG\,0043+039 was not only a temporal event and was not only caused by strong 
X-ray variations and or optical/UV variations
in this BAL quasar.
In a first paper (Kollatschny et al.\citealt{kollatschny15}, hereafter called
Paper I) we briefly presented our X-ray detection. Furthermore,
we discussed newly detected strong,
broad humps seen in the UV spectrum of PG\,0043+039
taken with the HST. 
We attributed these humps to
cyclotron lines.

\section{Observations}

We took simultaneous X-ray, UV, and optical spectra of PG\,0043+039 in July
2013:

\subsection{XMM-Newton observations}

PG\,0043+039 was observed twice with XMM-Newton (Jansen et al.\citealt{Jansen2001}).
The first observation (Obs1 in the following) was performed on the 15.6.2005 under ObsId. 0300890101.
This observation was free of any periods of high background radiation implying exposure times of 26.6~ks for pn (Str{\"u}der et al.\citealt{Strueder2001}),
31.2~ks for MOS~1 (Turner et al.\citealt{Turner2001}), and 31.1~ks for MOS~2.
The second observation (Obs2) was performed on the 18.7.2013 under ObsId. 0690830201.
This observation was affected by high background radiation periods.
The data were processed with SAS 14.0.0 in January 2015, using the latest calibration and following the methods described in Schartel et al.\cite{Schartel2007}
 and Piconcelli et al.\cite{Piconcelli2005}. 
All effective areas were calculated applying the method {\it corrarea,} which empirically corrects the EPIC effective areas for the instrumental differences. 
We screened for low background periods:
For the energy range from 0.2~keV to 12~keV, we extracted the counts that were registered within an annulus centered at the optical position of PG\,0043+039 with an inner radius of 1~arcmin and an outer radius of  11~arcmin for pn (14~arcmin for the MOSs).
To generate a light curve we binned the counts with 100~s.
We defined low background times as time intervals with a count rate below
6 c/s for pn and 4 c/s for MOSs.
We obtained a clean exposure time of 14.5~ks for pn, 29.0~ks for MOS1,
and 31.3~ks for MOS2.
For each modeling of the X-ray spectra, we assumed Galactic foreground
absorption with an column density of
 N$_H$~$=$~3.0~$\times$ 10$^{20}$~cm$^{-2}$ (Savage et al.\citealt{Savage1993}).

During Obs1 several observations
were performed with the OM (Mason et al.\citealt{Mason2001}). 
We took the source fluxes and their errors for the image observations with different filters from the second release of the XMM OM Serendipitous Ultraviolet Source Survey catalog (XMM-SUSS2).
The description of the first release of the catalog can be found in Page et al.\cite{Page2012}.
The spectrum from OM V-grism
% of exposure 14
 was manually extracted
 using omgsource, to avoid the close-by spectrum of a star that contaminates
the background of the target spectrum in the automatic extraction by SAS.
%The UV grism spectrum is not very useful as being very contaminated and noisy
%(WK: its not that bad!).
During Obs2 OM could not operate as PG~0043+039 was too close to Uranus
to enable  safe operation of the instrument. 
%However, the simultaneously taken spectra with HST and SALT more that
% compensate for this. 

\subsection{HST-COS FUV spectroscopy}

We observed the BAL PG\,0043+039 over one full HST orbit
at RA, Dec (J2000) = 00:45:47.230, +04:10:23.40
 with an exposure time 
of 1552 seconds on July 18 2013. 
We used the far-ultraviolet (FUV) detector of the Cosmic Origins Spectrograph (COS) with the G140L grating and an 2.5 arcsec
aperture (circular diameter).
 This spectral set covers the wavelength range from $\sim$ 1140\,\AA\
to $\sim$ 2150~\AA\
with a resolving power of 2000 at 1500\,\AA{}.
To fill up the wavelength hole produced by the chip gap and to reduce
the fixed pattern noise, we split our observation into four separate segments
of 388 s duration at two different FP-POS offset positions and four different
central wavelengths.
The observed spectrum corresponds to $\sim$ 800\,\AA{} to $\sim$ 1550~\AA\
in the rest frame of the galaxy.
 The original data were processed using the
standard CALCOS calibration pipeline. 
We corrected this UV spectrum, as well as our optical spectra of PG\,0043+039,
for Galactic extinction.
We used the reddening value E(B-V) = 0.02087 deduced from
 the Schlafly \& Finkbeiner\cite{schlafly11} recalibration of the 
Schlegel et al.\cite{schlegel98} infrared-based dust map. The
reddening law of Fitzpatrick\cite{fitzpatrick99} with R$_{V}$\,=\,3.1
was applied to our UV/optical spectra.

\subsection{Ground-based optical spectroscopy with the SALT
 and HET telescopes}

We took one optical spectrum of PG\,0043+039 with the 10m Southern African
Large Telescope (SALT)
nearly simultaneously with the XMM/HST observations on July 21, 2013
under photometric conditions.
However, the moon was bright during this observation. 
The spectrum was taken with the Robert Stobie Spectrograph
(RSS; see Burgh et al.\citealt{burgh03}) attached to the telescope
 using the pg0900
grating with a 1.5 arcsec wide slit. 
%under seeing conditions of 1.5 arcsec
With a grating angle of 21.125 degrees,  we covered the wavelength range from
6445 to 9400~\AA\  at a spectral resolution of 4.8~\AA\ (FWHM) and a reciprocal
dispersion of 0.97~\AA pixel$^{-1}$. The observed wavelength range corresponds
to a wavelength range from 4653 to 6786~\AA\ in the rest frame of the galaxy. 
 There are two gaps in the spectrum caused by the gaps between the three CCDs:
one between the blue and the central CCD chip as well as one between the
central and red CCD chip, covering the wavelength ranges
7425-7480~\AA\  and 8438-8491~\AA\ (5360-5400~\AA\ and 6092-6130~\AA\
in the rest frame).
The exposure time of our spectrum was 2200 seconds (37 minutes),
 which yielded a  S/N  of 118 at 7020~$\pm$10~\AA{} in the observed frame.

In addition to the galaxy spectrum, necessary flat-field and
Xe arc frames were observed, as well
as spectrophotometric standard stars for flux calibration (Hiltner~600, LTT4363).
The spectrophotometric standard stars
were used to correct the measured counts for the combined
transmission of the instrument, telescope, and atmosphere
as a function of wavelength.
Flat-field frames
were used to correct for differences in sensitivity both between
detector pixels and across the field.
The bright moon caused fringes in the red CCD at around 9000~\AA{}.

 We took a second optical spectrum of PG\,0043+039
 with the 9.2m Hobby-Eberly Telescope  (HET) at McDonald Observatory 
 on August 1, 2013 under nearly photometric conditions.
The spectrum was taken with the
Marcario Low Resolution Spectrograph (LRS)
mounted at the prime focus of HET. The detector was
a $3072\times1024$ 15 $\mu$m pixel Ford Aerospace CCD with 2x2 binning. 
This spectrum covers the wavelength range from 4390\,\AA\
to 7275~\AA\ (LRS grism 2 configuration)
with a resolving power of 650 at 5000\,\AA\ (8.2\,\AA\ FWHM).
This wavelength range corresponds to 3170 to 5250~\AA{}
in the rest frame of the galaxy.
The spectrum was taken with an exposure time of 
1500 seconds (25 minutes), which yielded a     
S/N of 102 at 5120~$\pm$10~\AA{} and of 83 at 7020~$\pm$10~\AA{}.
The slit width was
2\arcsec\hspace*{-1ex}.\hspace*{0.3ex}0 projected on the
sky. We took
Xe spectra  to enable the
wavelength calibration. A spectrum of the standard star BD40 was
observed for flux calibration as well.

Our SALT spectrum has a spatial
resolution of 0\arcsec\hspace*{-1ex}.\hspace*{0.3ex}2534 per binned pixel.
We extracted eleven columns from each of our object spectra 
corresponding to 2\arcsec\hspace*{-1ex}.\hspace*{0.3ex}8.
Our HET spectrum has a spatial
resolution of 0\arcsec\hspace*{-1ex}.\hspace*{0.3ex}472 per binned pixel.
%Furthermore, all observations were taken at the same airmass
%thanks to the particular design feature of the HET.
Here we extracted seven columns for our object spectrum 
corresponding to 3\arcsec\hspace*{-1ex}.\hspace*{0.3ex}3.

The reduction of the spectra (bias subtraction, cosmic ray correction,
flat-field correction, 2D-wavelength calibration, night sky subtraction, and
flux calibration) was performed in a homogeneous way with IRAF reduction
packages (e.g., Kollatschny et al.\citealt{kollatschny01}). 
We corrected the optical spectra for atmospheric absorption bands as well.
All wavelengths were converted to the rest frame of the galaxy (z=0.38512).  
%Throughout this paper, we assume that H$_0$~=~70~km s$^{-1}$ Mpc$^{-1}$.

\section{Results}

Here we present the results of our observing campaign based on the obtained
spectral data in the
 X-ray, UV, and optical frequency bands.

\subsection{X-ray flux in PG\,0043+039}

Visual inspection of the EPIC images of Obs1 does not reveal an X-ray counterpart for PG\,0043+039 and Czerny et al.\cite{czerny08} derived an upper limit for the source flux of  $<$8.6~$\times$~10$^{-16}$~ergs~s$^{-1}$~cm$^{-2}$
 for the 0.1 to 2.4 keV energy range.
The nearest X-ray source to PG\,0043+039 is 3XMM J004548.8+041018
(Watson et al.\citealt{Watson2009}) at a distance of 24.06~arcsec.
In 3XMM-DR5 the source has a EPIC (CR(pn) \& CR(MOS1) \& CR(MOS2))
 count rate of 1.92~$\pm$0.10~$\times$~10$^{-2}$counts~s$^{-1}$
in the 0.2 to 12. keV energy range
 and is most likely the counterpart to SDSS-DR8 J004548.80+041019.0.
In order to check for weakest X-ray signal of PG\,0043+039, we reanalyzed
 all three EPIC exposures together.
We extracted the possible source counts in a circle centered at the optical position of PG\,0043+039 with a radius of 10~arcsec for pn and 12~arcsec of MOS and we extracted the background counts from a circle centered at  0:45:43.62 +4:10:00.95 with a radius of 34~arcsec.
The background area is source free and located at the same CCD. 
The combined analysis leads to a weak signal with
a total EPIC pn+MOS count rate of
3.7~$\pm$~1.1~$\times$~10$^{-4}$counts~s$^{-1}$ for the 0.3 to 12.0~keV energy range.
We performed the same analysis for three control positions
(0:45:48.15 +4:10:41.79, 0:45:49.73 +4:10:40.53, and 0:45:50.53 +4:10:20.74).
Each position shows a distance of ~24~arcsec to 3XMM J004548.8+041018
and is located at the same CCD as PG\,0043+039.
We obtained the following count rates: 0.43~$\pm$~9.42~$\times$~10$^{-5}$~counts~s$^{-1}$,
 0.70~$\pm$~9.00~$\times$~10$^{-5}$~counts~s$^{-1}$, and 
-0.92~$\pm$9.19~$\times$~10$^{-5}$~counts~s$^{-1}$.
The obtained count rates of the control positions are significantly lower than the signal obtained for the optical position of PG\,0043+039.
The obtained signal cannot be explained with the short distance to 3XMM J004548.8+041018
 and we conclude that the combined analysis of all there EPIC cameras reveal a weak X-ray signal from PG\,0043+039

3XMM J004548.8+041018 is present in the Obs2, too, but the source shows
a significantly decreased flux with an Epic count rate of
4.19~$\pm$0.73~$\times$~10$^{-5}$~counts~s$^{-1}$ for the
0.2 to 12 keV region.
PG\,0043+039 is clearly visible as point source in the images of all there exposures.
We extracted source counts exactly as described for Obs1 above except that we centered the circle at the eye-determined center of the X-ray emission (0:45:47.07 +4:10:23.50) and obtained a count rate of 1.42~$\pm$0.17~$\times$~10$^{-3}$~counts~s$^{-1}$ for all EPIC cameras together, in the 0.3 to 12 keV energy band.
We extracted the background counts from a circle centered at 0:45:43.41 +4:10:04.25 
with a radius of 34~arcsec.
We repeated the analysis for three control positions (0:45:48.03 +4:10:42.07, 0:45:49.66 +4:10:40.22, 0:45:50.43 +4:10:16.87) selected as described above and obtained count rates
that are in agreement with no source flux -0.63~$\pm$~0.93~$\times$~10$^{-4}$~counts~s$^{-1}$, -1.17~$\pm$~0.88~$\times$~10$^{-4}$~counts~s$^{-1}$, and 
-1.40~$\pm$0.87~$\times$~10$^{-4}$~counts~s$^{-1}$.

\subsection{X-ray spectra of PG\,0043+039}

Given the low number of accumulated counts we added the pn, MOS1, and MOS2 spectra for each observation and calculated the corresponding auxiliary files. 
For the calculations, each input file was weighted with the corresponding exposure time of the camera.
All modeling was done with {\it Xspec 12.8.2} (Arnaud, \citealt{Arnaud1996}).
We used C-statistics and a spectrum that was slightly binned, such that each bin contains at the minimum two counts. 
In addition, for Obs. 2, we analyzed  a spectrum that was binned such that each bin contained 15 counts. 
For this spectrum, we applied the $\chi^2$-statistics and F-test.
   \begin{figure*}
\centering
    \includegraphics[width=8.1cm,angle=270]{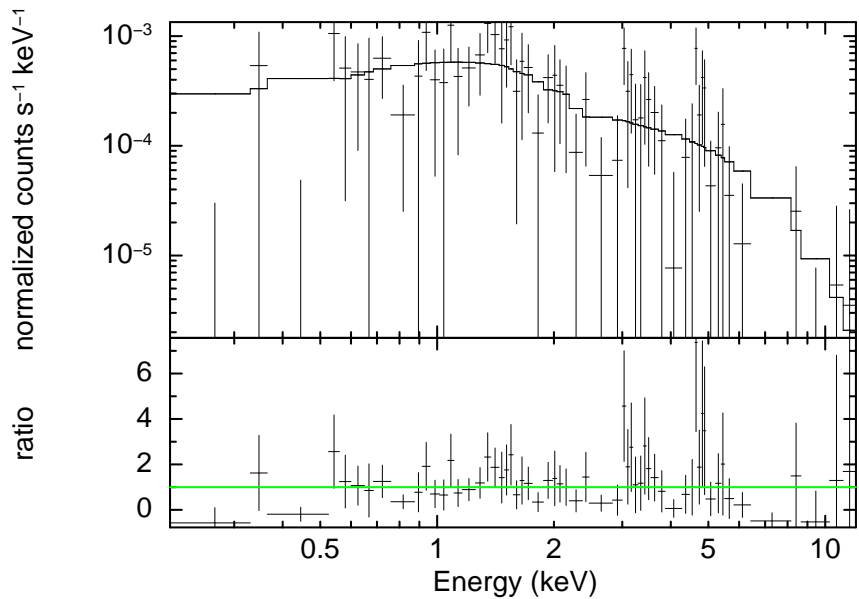}
      \caption{EPIC spectrum of PG\,0043+039 from 2013 is shown in comparison to the best-fit power law absorbed by Galactic column density.
 Data are slightly binned, such that each bin contains at the minimum, two source counts. 
              }
       \vspace*{-3mm} 
         \label{xspec_nice_plot_30358.ps2}
   \end{figure*}
   \begin{figure*}
\centering
    \includegraphics[width=8.1cm,angle=270]{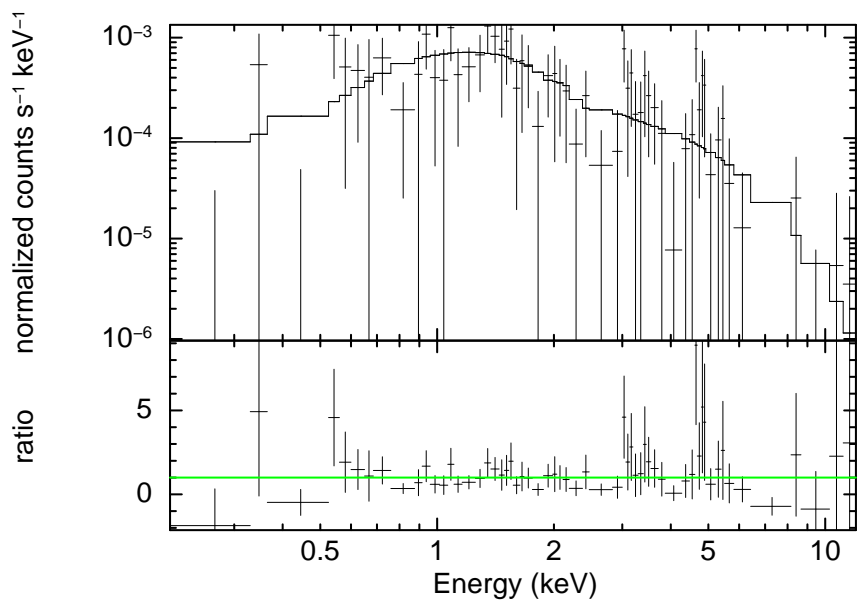}
      \caption{EPIC spectrum of PG\,0043+039 from 2013 is shown in comparison to the best-fit power law absorbed by Galactic column density and intrinsic neutral absorption. Data are slightly binned, such that each bin contains, at the minimum, two source counts. 
              }
       \vspace*{-3mm} 
         \label{xspec_nice_plot_31439.ps2}
   \end{figure*}

   \begin{figure*}
\centering
    \includegraphics[width=8.1cm,angle=270]{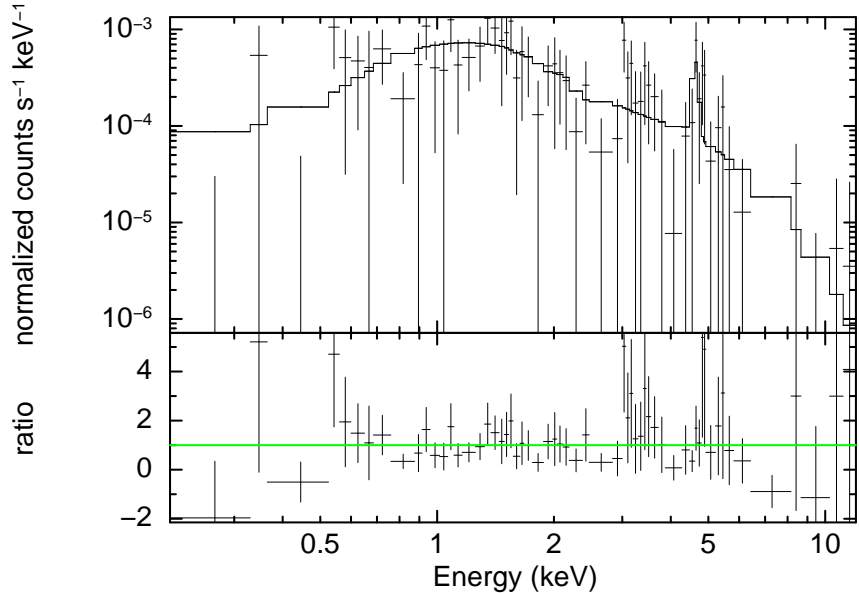}
      \caption{EPIC spectrum of PG\,0043+039 from 2013 is shown in comparison to
 the best-fit intrinsic absorbed power law plus neutral iron line.
Data are slightly binned, such that each bin contains, at the minimum, two
 source counts.
              }
       \vspace*{-3mm} 
         \label{xspec_nice_plot_32282.ps2}
   \end{figure*}
   \begin{figure*}
\centering
    \includegraphics[width=8.1cm,angle=270]{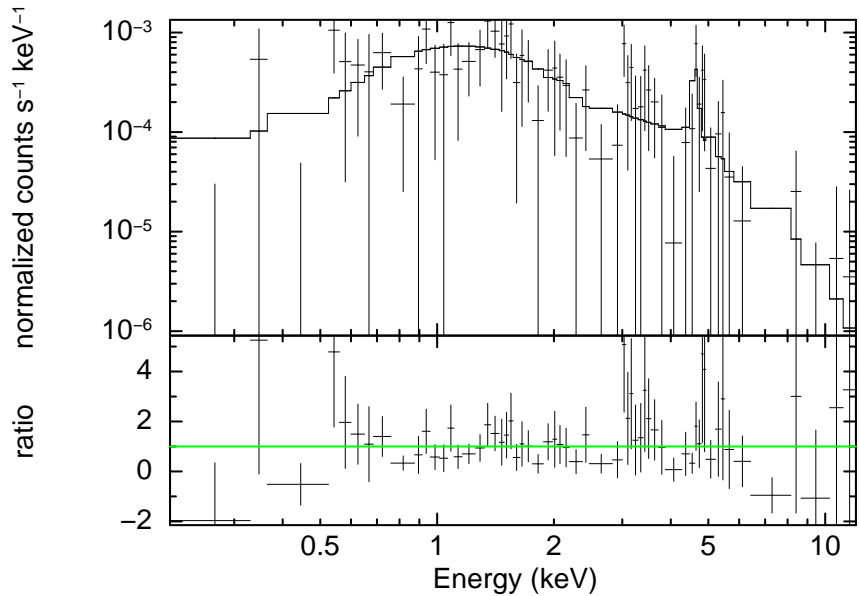}
      \caption{EPIC spectrum of PG\,0043+039 from 2013 is shown in comparison to
 model assuming an absorbed primary power-law continuum and reflection on
 distant, optical thick material. 
Data are slightly binned, such that each bin contains, at the minimum, two
 source counts. 
              }
       \vspace*{-3mm} 
         \label{xspec_nice_plot_13288.ps2}
   \end{figure*}
The errors of best-fit parameters are provided to the 90\% confidence level if not stated otherwise. 
For the description of absorption, we used the xspec model {\it tbnew,} which is based on Wilms et al.\cite{Wilms2000}, solar abundances from Wilms et al.\cite{Wilms2000}, and photoelectric cross-sections from Verner et al.\cite{Verner1996}.
%For each modeling we assumed Galactic foreground absorption with an column density of  N$_H$~$=$~3.0~$\times$ 10$^{20}$~cm$^{-2}$ (Savage et al.\citealt{Savage1993}).

\paragraph{Observation 2:} %\\

\noindent
Figure~\ref{xspec_nice_plot_30358.ps2} shows the best-fit power law on the EPIC spectrum 
 of PG\,0043+039 from 2013 compared to the data. 
The fit (C~$=$~61.5, d.o.f.~$=$~58) reveals a very hard spectrum with $\Gamma$~$=$~1.09$\pm$0.24.
We therefore modeled the data with a power law absorbed by  neutral material at the redshift of the quasar (Fig.~\ref{xspec_nice_plot_31439.ps2}),  which  allowed us to decrease C by $\Delta$C~$=$~6.0.
The residuals show enhanced emission at about 4.8~keV, which could correspond to neutral iron K$_{\alpha}$ in the rest frame of the quasar. 
We therefore added a neutral iron line to the model (E~$=$~6.4~keV, $\sigma$~$=$~10~eV),
 which allowed us to further decrease C by $\Delta$C~$=$~5.8
(Fig.~\ref{xspec_nice_plot_32282.ps2}).
For the fit, we obtained C~$=$~49.7 for d.o.f.~$=$~56
with the following parameters:
N$_H$~$=$~$5.5_{-3.9}^{+6.9}$~$\times$10$^{21}$~cm$^{-2}$, 
$\Gamma$~$=$~1.70$_{-0.45}^{+0.57}$,
N(power law)~$=$~$6.6_{-2.9}^{+6.8}$~$\times$10$^{-6}$~keV$^{-1}$~cm$^{-2}$~s$^{-1}$ at 1 keV and 
N(Gauss)~$=$~$3.8_{-2.8}^{+4.1}$~$\times$10$^{-7}$~photons~cm$^{-2}$~s$^{-1}$.
For the total model, we obtained the following fluxes: 
F(2.0-10.0~keV)~$=$~1.80$_{-0.29}^{+0.24}$~$\times$10$^{-14}$~ergs~cm$^{2}$~s$^{-1}$,
F(0.2-2.0~keV)~$=$~5.22$_{-0.95}^{+0.42}$~$\times$10$^{-15}$~ergs~cm$^{2}$~s$^{-1}$, and
F(0.2-12.0~keV)~$=$~2.48$_{-0.38}^{+0.18}$~$\times$10$^{-14}$~ergs~cm$^{2}$~s$^{-1}$, where the flux errors are provided for the 68\% confidence.

In addition we modeled the EPIC spectrum of Obs.~2, binned such that each bin containing 15 counts, using $\chi^2$-statistics. 
We obtained $\chi^2$~$=$~12.9 at d.o.f.~$=$~8 for a simple power-law fit
% (NS: fit 8/1, xspec\_plot\_15143.pdf) 
and  $\chi^2$~$=$~4.6 at d.o.f.~$=$~6 
%  (NS: fit 8/2, xspec\_plot\_15229.pdf)
for an intrinsically absorbed power law plus an iron line. 
An F-test shows that the probability of finding the improvement by random chance
 is below 5\%.

The performed fits applying the C-statistics show that allowing intrinsic
 absorption and then adding an iron line both  improve the description
 of the data.
The achieved decrease of C does not allow us to claim a significant detection.
The F-test confirms that  modeling the data with intrinsic absorption
 and an iron line
improves the description, but does not allow us to claim a significant detection.

X-ray weakness of quasars might be explained with a completely absorbed X-ray
 continuum in combination with a weak so-called escaping reflection component.
Therefore, we modeled the EPIC spectrum of Obs.~2, assuming an absorbed primary
 power-law continuum and an unabsorbed reflection component.
We described the reflection with the model {\it pexmon,} which is provided in
 {\it Xspec} (Nandra et al.\citealt{Nandra2007}). 
This model assumes neutral Compton reflection at distant, optically thick
 material and considers line emission consistently
(Nandra et al.\citealt{Nandra2007}). 

Figure~\ref{xspec_nice_plot_13288.ps2} shows the best-fit model 
%(NS: fit 3/3, xspec\_plot\_13288.pdf) 
assuming an intrinsic absorbed power law plus 
 an unabsorbed reflection compared to the data. 
For the fit, we obtained C~$=$~49.6 for d.o.f.~$=$~56.
The primary power-law continuum is absorbed with 
N$_H$~$=$~6.3$_{-3.2}^{+5.0}$~$\times$10$^{21}$~cm$^{-2}$. 
As the power-law index is poorly constrained, we fixed the index to $\Gamma$~$=$~1.9, which corresponds to the mean value obtained for radio-quiet quasars of the  Palomar-Green (PG) Bright Quasar Survey sample for the 2keV to 10 keV energy range    (Piconcelli et al.\citealt{Piconcelli2005}) and obtained 
N(power law)~$=$~7.4$_{-2.2}^{+2.8}$~$\times$10$^{-6}$~keV$^{-1}$~cm$^{-1}$~s$^{-1}$. 
For the reflection component, we fixed the photon index to $\Gamma$~$=$~1.9
 (Piconcelli et al.\citealt{Piconcelli2005}), the cut-off energy to 100~keV, 
the metals to solar abundance, and the inclination angle to 45 degree. 
We determined the iron abundance Fe$_{abund.}$~$=$~6.8$_{-5.9}^{+94.}$ and  
N({\it pexmon})~$=$~$2.0_{-1.5}^{+2.0}$~$\times$10$^{-5}$~photons~cm$^{-2}$~s$^{-1}$ at 1~keV; see Nandra et al.\cite{Nandra2007} and Arnaud\cite{Arnaud1996} for further details. 
%%%%%%%%%%%%%%%%%%%%%%%%%%%%%%%%%%%%%%%%%%%%%%
   \begin{figure*}
\centering
    \includegraphics[width=10cm,angle=270]{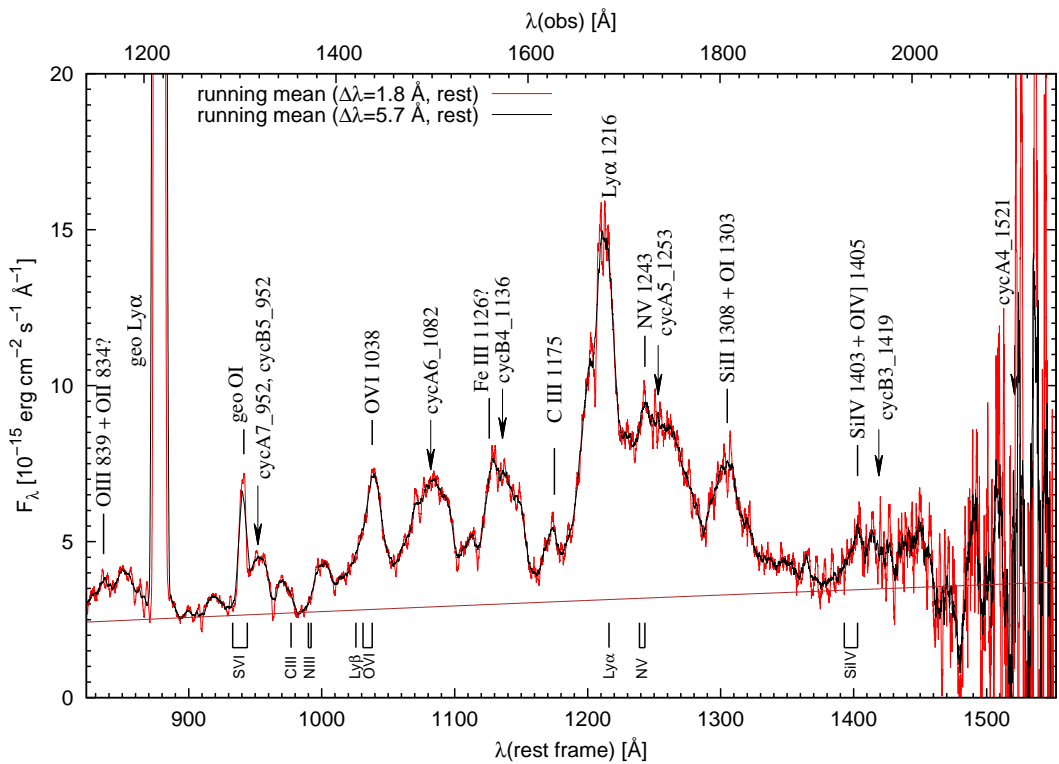}
      \caption{HST-COS FUV spectrum of PG\,0043+039 corrected for Galactic
       reddening. The brown line (see Paper 1)
            shows an approximation of the continuum. Below the spectrum
  the locations of possible absorption lines are indicated.    
              }
%       \vspace*{-3mm} 
         \label{pg0043_hst.ps}
   \end{figure*}

%%%%%%%%%%%%%%%%%%%%%%%%%%%%%%%%%%%%%%%%%%%%%%
Finally, we tested whether an absorbed, reflection-only model could describe the data.
The idea is to assume a completely absorbed primary continuum so that only the reflected component can reach an observer in our direction.  
The obtained C values were significantly larger with $\Delta$~C~$=$~26.1 for d.o.f.~$=$~55.
% (NS: fit 2/2, xspec\_plot\_6166.pdf). 
In addition, we tried to fit the spectrum, which is binned with 15 counts per bin, with this model by applying the $\chi^2$-statistics.
%(NS: fit 2/6, xspec\_plot\_10692.pdf ).
We were unable to obtain a statistically valid description of the data and therefore we exclude this scenario.

\paragraph{Observation 1:} %\\

\noindent 
Given the low number of counts collected during the XMM-Newton observation of 
PG\,0043+039 in 2005 we analyzed the data compared to our findings described
in the previous paragraph.
This analysis was led by two questions: (1) Can we detect the components that
 we found in Obs~2? and (2) Are the components absorbed by additional material that was not present in Obs~2?

We fitted the EPIC spectra of Obs.1 with a model consisting of an absorbed
power law plus iron emission line. 
With exception of the normalizations, all parameters were fixed to the  values
obtained for Obs.~2, e.g.,  
N$_H$~$=$~5.5$\times$10$^{21}$~cm$^{-2}$ and $\Gamma$~$=$~1.70.
We
% (NS fit 7/1 xspec\_plot\_11125.pdf)
 obtained C~$=$~66.3 for d.o.f.~$=$~40 with 
N(power law)~$=$~1.33$_{-0.58}^{+0.69}$~$\times$~10$^{-6}$~keV$^{-1}$~cm$^{-2}$~s$^{-1}$ at 1 keV and 
N(Gauss)~$=$~1.9$_{-1.9}^{+1.9}$~$\times$~10$^{-8}$~photons~cm$^{-2}$~s$^{-1}$.
Given these numbers we are not able to detect the iron line in Obs~1.
We then took the power-law continuum as determined for Obs~2 and determined
a formal absorbing column density, which is required to model the data of Obs~1.
We obtained
% (NS: fit 7/2 xspec\_plot\_11358.pdf)
N$_H$(power law)~$=$~5.4$_{-3.3}^{+6.9}$~$\times$~10$^{23}$~cm$^{-2}$ 
 (C~$=$~75.5 for d.o.f.~$=$~41).
We applied the same procedure for the Gaussian line
%(NS: 7/3 xspec\_plot\_11486.pdf)
 fit and obtained: 
N$_H$(Gauss)~$=$~1.7$_{-1.4}^{+1.5}$~$\times$~10$^{24}$~cm$^{-2}$
(C~$=$~83.7 for d.o.f.~$=$~41).

We followed the same strategy with the physical model assuming an absorbed
 primary power-law continuum and Compton reflection on distant, optically
thick material (Nandra et al.\citealt{Nandra2007}).
Again we fixed all model parameters to the best-fit values obtained for Obs.~2
with the exception of the normalizations, which were free to vary.
We obtained
% (NS fit 7/5 xspec\_plot\_12259.pdf)
 C~$=$~66.7 for d.o.f.~$=$~40 with 
N(power law)~$=$~1.50$_{-0.72}^{+0.85}$~$\times$~10$^{-6}$~keV$^{-1}$~cm$^{-2}$~s$^{-1}$ at 1 keV and 
N(reflection)~$=$~2.1$_{-2.1}^{+9.4}$~$\times$~10$^{-6}$~photons~cm$^{-2}$~s$^{-1}$.
In agreement with the analysis above, we conclude that we are unable
 to detect the reflection component in the data of Obs.~1.
Similarly, we determined formal column densities for the physical model. 
For each component (continuum and reflection) individually we froze the model parameters to the values of the best fit of Obs.~2 and determined
a formal column density required to describe the data of Obs.~1.
We obtained
% (NS: fit 7/6 xspec\_plot\_12405.ps2)
N$_H$(power law)~$=$~4.3$_{-3.6}^{+7.0}$~$\times$~10$^{23}$~cm$^{-2}$ 
 (C~$=$~76.3 for d.o.f.~$=$~41) and
%  (NS: 7/7 xspec\_plot\_12554.ps2)  
N$_H$(reflection)~$=$~1.1$_{-1.1}^{+5.0}$~$\times$~10$^{24}$~cm$^{-2}$
(C~$=$~82.4 for d.o.f.~$=$~41).
%%%%%%%%%%%%%%%%%%%%%%%%%%%%%%%%%%%%%%%%%%%%

\subsection{HST UV/FUV spectra of PG\,0043+039}
%\subsubsection{UV/FUV spectrum PG\,0043+039}

The UV spectrum of PG\,0043+039 we took with
HST in 2013 is shown in
Fig.~\ref{pg0043_hst.ps}.
The observed wavelength range from $\sim$1140\,\AA{}
to $\sim$2150~\AA{} corresponds to an intrinsic wavelength range of
$\sim$820\,\AA{} to $\sim$1550~\AA{}.
   \begin{figure*}
\centering
    \includegraphics[width=10cm,angle=270]{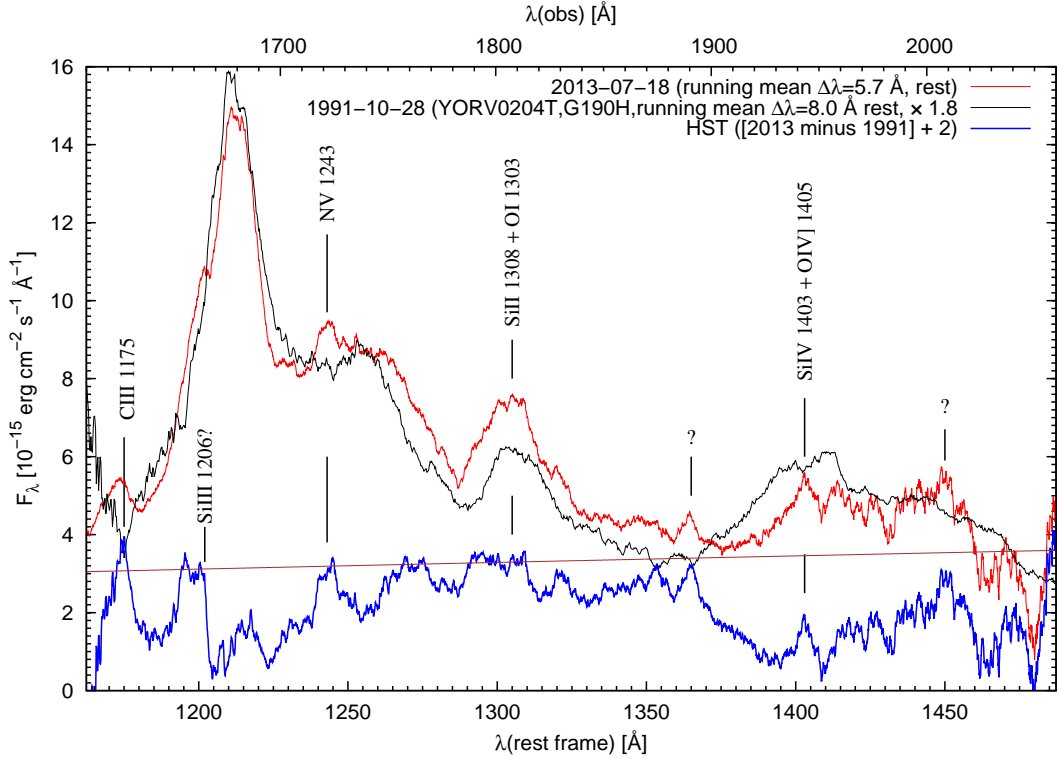}
      \caption{UV spectral variability: HST UV spectra of PG\,0043+039
 taken in the
 years 2013 (red)
 and in 1991, with flux multiplied by a factor 1.8 (black), as
 well as their difference spectrum (blue).
              }
%       \vspace*{-3mm} 
         \label{pg0043_hst_2013_1991.ps}
   \end{figure*}
   \begin{figure*}
\centering
    \includegraphics[width=10cm,angle=270]{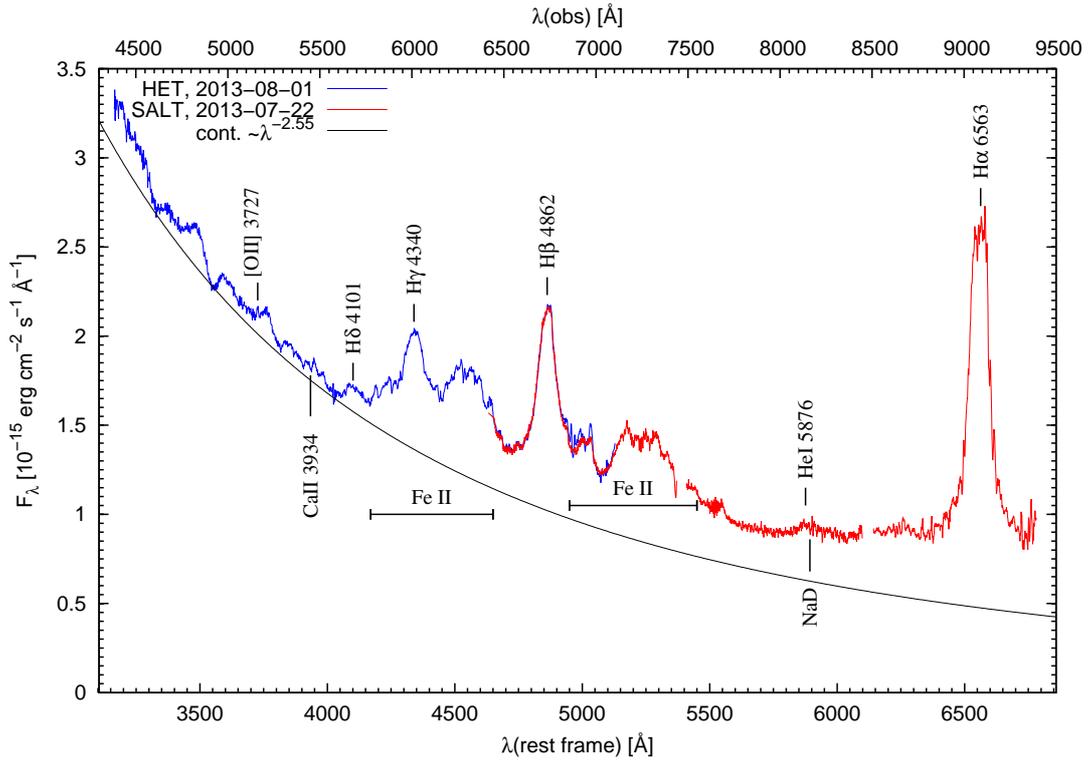}
      \caption{Combined optical spectrum of PG\,0043+039 taken with the
 HET and SALT telescopes in 2013.
              }
 %      \vspace*{-3mm} 
         \label{pg0043_het_salt.ps}
   \end{figure*}

\begin{table*}
%\centering
\tabcolsep+2.8mm
\caption{Emission line intensities in rest frame and corrected for Galactic extinction}
\begin{tabular}{lcccccl}
\hline 
\noalign{\smallskip}
Emission line                 &Flux     & Rel.Flux  &    Wavelength range & \multicolumn{2}{c}{Pseudo-continuum}      &Telescope \\
                              &         &           &        [\AA{}]      & blue side [\AA{}]  &    red side  [\AA{}] & \\
\noalign{\smallskip}
(1)                           & (2)     &  (3)      &    (4)              &        (5)  &         (6)                 &  (7)\\
\noalign{\smallskip}
\hline 
\noalign{\smallskip}
??                            & 53.3$\pm{}$4.    & .682      &   812 --  869   &               &               & HST\\
cycB5,A7                    & 50.0$\pm{}$4.    & .640      &   927 --  981   &               &               & HST\\
??                            & 24.2$\pm{}$3.    & .310      &   984 -- 1011   &               &               & HST\\
$\ion{O}{vi}\,\lambda 1038$   & 98.9$\pm{}$7.    & 1.27      &  1011 -- 1054   &               &               & HST\\
cycA6                       & 144$\pm{}$11.    & 1.84      &  1054 -- 1102   &               &               & HST\\
cycB4                       & 137$\pm{}$11.    & 1.75      &  1118 -- 1161   &               &               & HST\\
$\ion{C}{iii}\,\lambda 1175$  & 33.2$\pm{}$11.   & .425      &  1162 -- 1081   &               &               & HST\\
Ly$\alpha$                    & 339$\pm{}$22.    & 4.34      &  1181 -- 1234   &               &               & HST\\
$\ion{N}{v}\,\lambda 1243$    & 10.8$\pm{}$7.    & .138      &  1234 -- 1248   &  1709         &               & HST\\
cycA5                       & 202$\pm{}$22.    & 2.59      &  1243 -- 1287   &               &               & HST\\
$\ion{Si}{ii}\,\lambda 1306 + \ion{O}{i}\,\lambda 1303$ & 103$\pm{}$14     &  1.32    &  1287 -- 1318   &               &               & HST\\
$\ion{Si}{v}\,\lambda 1403$ + cycB3 &93.9$\pm{}$22.&1.20   &  1375 -- 1460   &              &               & HST\\
$[\ion{O}{ii}]\,\lambda 3727$ &   0.33$\pm{}$ .04 & .004 & 3720 -- 3732 & 3719 -- 3721 &  3731 -- 3733 & HET\\
H$\delta$                     &  3.6$\pm{}$ 1.0  & .046 & 4060 -- 4155  & 4040 -- 4060 &  4690 -- 4745 & HET\\
H$\gamma$                     &  22.9$\pm{}$ 4.0 & 0.29 & 4265 -- 4435 & 4255 -- 4265  &  4435 -- 4445 & HET\\
\ion{Fe}{ii}\,$\lambda\lambda 4500$  &  98.1$\pm{}$ 4.0 & 1.26 & 4155 -- 4690  & 4040 -- 4060 &  4690 -- 4745 & HET\\
H$\beta$                      &  78.1$\pm{}$ 4.0   & 1.00 & 4745 -- 4965  & 4690 -- 4745 &  5635 -- 5755 & HET/SALT\\
\ion{Fe}{ii}\,$\lambda\lambda 5020$  &  13.9$\pm{}$ 1.0 & 0.18  & 4965 -- 5070 & 4690 -- 4745 &  5635 -- 5755 & SALT\\
\ion{Fe}{ii}\,$\lambda\lambda 5320$  &  83.2$\pm{}$ 4.0 & 1.07 & 5070 -- 5635 & 4690 -- 4745 &  5635 -- 5755 & SALT\\
\ion{He}{i}\,$\lambda 5876$   &  5.3$\pm{}$ 1.0 & .068& 5790 -- 6005 & 5745 -- 5790 &  6005 -- 6045 & SALT\\
H$\alpha$                     &  186.$\pm{}$ 15. & 2.38 & 6360 -- 6715 & 6310 -- 6360  &  6715 -- 6750 & SALT\\
%H$\beta$                     & HET   &  77.0$\pm{}$   & 4745\,\AA{} -- 4965\,\AA{}  & 4690\,\AA{} -- 4745\,\AA{}  &  5060\,\AA{} -- 5080\,\AA{}\\
%\noalign{\smallskip}
%Ly$\alpha$                    & HST   &  263.1$\pm{}$  & 1180\,\AA{} -- 1233\,\AA{}  &     \,\AA{} --     \,\AA{}  &      \,\AA{} --     \,\AA{}\\
%Ly$\alpha$                    & HST   &  151.1$\pm{}$  & 1193\,\AA{} -- 1233\,\AA{}  &     \,\AA{} --     \,\AA{}  &      \,\AA{} --     \,\AA{}\\
%\ion{O}{vi}\,$\lambda 1038$   & HST   &  41.8$\pm{}$   & 1021\,\AA{} -- 1053\,\AA{}  &     \,\AA{} --     \,\AA{}  &      \,\AA{} --     \,\AA{}\\
%\noalign{\smallskip}
\noalign{\smallskip}
\hline
\end{tabular}\\
%Continuum flux () in units of 10$^{-15}$\,erg\,s$^{-1}$\,cm$^{-2}$\,\AA$^{-1}$.\\
Line fluxes (2) in units of 10$^{-15}$\,erg\,s$^{-1}$\,cm$^{-2}$.
\end{table*}

The HST-COS spectrum
% shown in Fig.~\ref{pg0043_hst.ps} 
 has been smoothed by means of
 two different running mean widths
($\Delta \lambda = 1.8$ and $5.7$\,\AA{} in rest frame) for highlighting
weaker spectral structures.  
We indicate the identifications of the strongest UV emission lines,
of the geo-coronal lines Ly$\alpha$ and $\ion{O}{vi}\,\lambda 1038$,
as well as of other emission lines, which we attribute
to two
cyclotron systems A and B with
their 3rd, 4th, 5th, 6th, and 7th harmonics. The integer numbers identify the
emission humps with multiples of the cyclotron fundamental
(see Paper I). This means, for example, that cycA5 is the fifth harmonic of the
cyclotron system A. The second number gives the wavelength.
We  discuss the relative intensities and the equivalent widths
of the strongest emission lines $\ion{O}{vi}\,\lambda 1038$, Ly$\alpha,$ 
and $\ion{N}{v}\,\lambda 1243$ in PG\,0043+039
in comparison to non-BAL quasars in Sect. 4.2.
We show a power-law continuum N$_\nu\sim\nu^{\alpha}$ with
$\alpha = 0.69 \pm 0.02.$ based on the near and far-UV spectra
(see  Fig.~\ref{pg0043_sed_dered_log.ps}
 in the discussion section).
The locations of possible absorption lines are indicated
below the spectrum.
We show the positions of the strongest absorption lines of an average
BAL quasar (taken from Baskin et al.\citealt{baskin13}, their Fig.~3).
The distribution of these possible absorption lines blueward of Ly$\alpha$  
cannot produce artificial emission features that are comparable with
the observed spectral humps.

During the first inspection of  our only FUV spectrum 
 (Fig.~\ref{pg0043_hst.ps}) it was difficult
to derive the UV continuum. However, regions in the long-ultraviolet spectrum 
at about    
1330 - 1350 , 1700 - 1720, and 1975 - 2000 \AA{} are generally free
from strong absorption or emission features in AGN
(e.g., Gibson et al.\citealt{gibson09}).
Therefore, we combined our short-wavelength UV spectrum with the long-wavelength
UV spectrum taken in 1991 and tried to fit a continuum in the
long-wavelength region first. 
Afterward, we extrapolated this continuum (a simple power law) to our
short-wavelength UV spectrum. This power law perfectly fits the
continuum in the 900 - 1000 \AA{} range.
At the end,  
we used the spectral ranges at 895, 983, 1360, 1610, 1690, and 1980\AA{}
with typical widths of 10\AA{} (see Paper I, Fig.~1) to fit the
UV continuum. 

The obtained UV/FUV spectrum of PG\,0043+039 is very exceptional in comparison to
other quasars and/or BAL quasars
(Hall et al.\citealt{hall02}; Baskin et al.\citealt{baskin13}; 
Saez et al.\citealt{saez12}) regarding the emission lines.
The intensities and the wavelengths of identified optical/UV lines
are given in Table 1. The relative line fluxes with respect to
H$\beta$~=~1.0 are given in column~3.  
There is no detectable Lyman edge associated with the BAL absorbing gas.

Figure~\ref{pg0043_hst_2013_1991.ps}
presents the common wavelength range of our HST-COS UV spectrum
taken in 2013 together with the
HST-FOS spectrum of PG\,0043+03 taken in 1991
(Turnshek et al.\citealt{turnshek94};
  multiplied by a factor of 1.8).
At the bottom of this figure the difference between the two spectra is shown
(i.e., the spectrum taken in 2013 subtracted by the spectrum taken in 1991).
Here some additional weak broad emission lines 
(e.g., $\ion{C}{iii}\,\lambda 1175$,
$\ion{N}{v}\,\lambda 1243$, $\ion{Si}{v}\,\lambda 1403$)  
sitting on top of broad bumps can be identified by
comparing our recent UV spectrum  with the HST-FOS spectrum taken in 1991
(see the discussion section). It is remarkable that these lines
were not identifiable in the UV spectrum taken in 1991.

\subsection{Optical spectra of PG\,0043+039}

A combined optical spectrum of PG\,0043+039 is shown in
Fig.~\ref{pg0043_het_salt.ps}
composed of the two spectra taken with the
 HET and SALT telescopes in 2013.

   \begin{figure*}
\centering
    \includegraphics[width=10cm,angle=270]{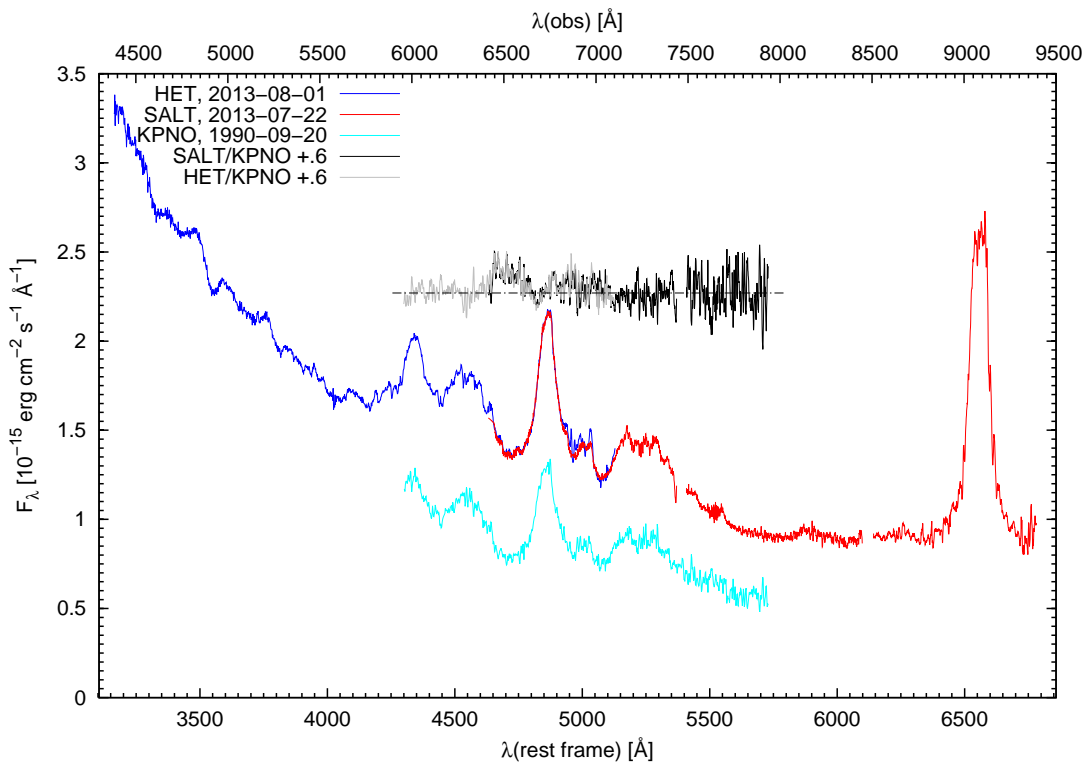}
      \caption{Optical variability: combined optical spectra of PG\,0043+039
 taken with the
 HET and SALT telescopes in 2013, an optical spectrum taken with the KPNO
in 1990, as well as the difference
between the spectra taken in 2013 and 1990 (additionally shifted by 
0.6). A horizontal line is added to the difference spectrum
to guide the eye.
              }
 %      \vspace*{-3mm} 
         \label{pg0043_het_salt_kpno.ps}
   \end{figure*}
\begin{figure*}
\centering
    \includegraphics[width=13cm,angle=0]{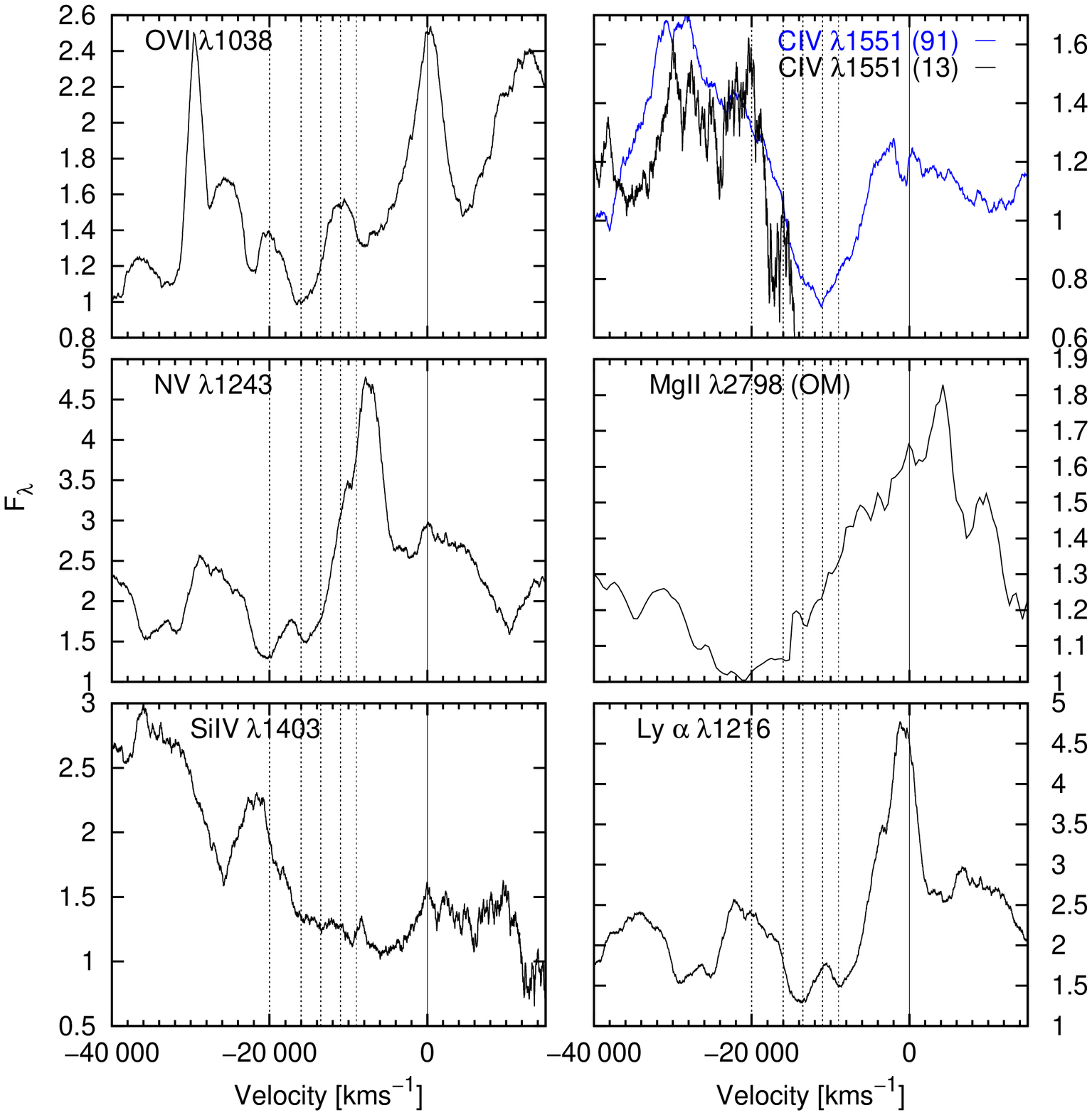}
      \caption{Possible absorption profiles belonging to the OVI,  NV, 
        SiIV, CIV, MgII, and  Ly$\alpha$ 
       lines. The profiles are divided  by the assumed continuum. The
 velocity scale is calculated
relative to the reddest line of each multiplet. The location of possible
 absorption lines, based on local minima, is
 indicated with the vertical tick marks at different velocities.} 
       \vspace*{-3mm} 
         \label{pg0043_velo3x2.ps}
   \end{figure*}

This  spectrum has been corrected for Galactic extinction.
The blue spectrum taken with the
HET telescope and the red spectrum taken with the SALT Telescope
perfectly overlap in the common H$\beta$ region. The observed total spectral
range from
 $\sim$4390\,\AA{} to $\sim$9400~\AA{} corresponds to a rest-frame spectrum from
$\sim$3170\,\AA{} to $\sim$6790~\AA{}.
PG\,0043+039 is a strong optical FeII emitter following Turnshek et al.\cite{turnshek94}. In addition,  Balmer lines
dominate the optical spectrum, and there are no indications of the presence of  narrow
$[\ion{O}{iii}]\,\lambda 5007, \lambda 4959$ lines.
In contrast
to Turnshek et al.\cite{turnshek94}, we clearly detect
in our recent spectra
a broad \ion{He}{i}\,$\lambda 5876$ line and
a weak narrow
$[\ion{O}{ii}]\,\lambda 3727$ line.
The underlying continuum of the
 optical spectrum shows a strong blue gradient.
A power-law continuum N$_\nu\sim\nu^{\alpha}$ with
$\alpha = -2.55 \pm 0.02$  is indicated based on 
 continua points at 3325, 3550, 3805, and 4010 \AA{}
with typical widths of 10\AA{}. 
We present in Table~1 the measured UV and optical emission line intensities
for PG\,0043+039 in rest frame and corrected for Galactic extinction.
We integrated the emission line intensities between the wavelength boundaries 
given in this table. First we subtracted a linear pseudocontinuum
defined by the wavelength ranges (Col. 5,6), then we integrated the emission
line flux. In some cases, we only extrapolated the continuum from one side. 

Figure~\ref{pg0043_het_salt_kpno.ps}
 shows the combined optical spectra of PG\,0043+039 taken with the
 HET and SALT telescopes in 2013 together with
 an optical spectrum taken before with the KPNO 2.1m telescope in Sept. 1990
(Boroson \& Green\citealt{boroson92}).
 The spectrum taken in 2013 is brighter
by a constant factor of 1.8 compared  to the spectrum taken in 1990.
This variability factor is similar to that of the UV spectra. 
In addition, the quotient spectrum 
 between these two epochs 
 (additionally shifted by 0.6) is given in Figure~\ref{pg0043_het_salt_kpno.ps}.
We added a horizontal line to guide the eye.
There are no clear spectral differences to be seen for these
two epochs. However, an extended blue wing in H$\beta$
that can be recognized in the FeII
subtracted spectrum of Boroson \& Green\cite{boroson92} might have varied.
 
\subsection{UV/opt. emission line profiles in PG\,0043+039}

\begin{table}
%\centering
%\tabcolsep+3.5mm
\tabcolsep+1.3mm
\caption{Line widths (FWHM) and shifts (uppermost 10 percent) of the strongest
 broad emission lines.}
\begin{tabular}{llcl}
\hline 
\noalign{\smallskip}
Emission Lines                &Width  &Shift &Telescope\\ 
                              &[\kms] &[\kms] &\\
\noalign{\smallskip}
(1)                           & (2)   & (3) & (4)\\
\noalign{\smallskip}
\hline 
\noalign{\smallskip}
\ion{O}{vi}\,$\lambda 1038$  &  4150 $\pm{}$ 200 & +390 $\pm{}$ 150 & HST\\
Ly$\alpha$                   &  6300 $\pm{}$ 500 & -860 $\pm{}$ 150 & HST\\
H$\beta$                     &  4750 $\pm{}$ 200 & 0                & SALT/HET\\
H$\alpha$                    &  4010 $\pm{}$ 200 & 0                & SALT\\
\noalign{\smallskip}
\hline
\noalign{\smallskip}
cycA6\,$\lambda 1082$        & 10\,200 $\pm{}$ 500 & 0 & HST\\
cycB4\,$\lambda 1136$        &  8600 $\pm{}$ 500 & 0 & HST\\
cycA5\,$\lambda 1253$        & 12\,700 $\pm{}$ 500 & 0 & HST\\
\noalign{\smallskip}
\hline
\end{tabular}\\
\end{table}

The emission line profiles contain information about the kinematics
of their line emitting regions. The strongest single emission lines
in the optical
spectra of PG\,0043+039 are the Balmer lines H$\alpha$ and H$\beta$
(see Fig.~\ref{pg0043_het_salt.ps}). Besides that, the optical spectrum
is dominated by FeII blends.   
The strongest single emission lines in the UV are the
Ly$\alpha$ and \ion{O}{vi}\,$\lambda 1038$ lines
(see  Fig.~\ref{pg0043_hst.ps}). There are further line identifications in our
FUV HST-COS spectrum, which we attribute to cyclotron lines. Theses lines
have different profiles.
We showed the optical and UV profiles  in velocity space in
Paper I.
%Fig.~\ref{pg0043_velo_profile_lya855_cont2.ps}.
%They are normalized to the same maximum intensity.
%
%\begin{figure}
%    \includegraphics[width=8cm,angle=0]{pg0043_velo_profile_lya855_cont2.ps}
%      \caption{Normalized emission line profiles of the
% strongest optical and UV emission lines in velocity space. The Ly$\alpha$ line
% has been shifted by 855 [\kms] to the red for comparing the profiles.} 
%       \vspace*{-3mm} 
%         \label{pg0043_velo_profile_lya855_cont2.ps}
%   \end{figure}
%
We present in Table~2 the line widths, i.e., the full width at half maximum
(FWHM) and the shifts of their emission line centers (centroid of the
uppermost 10 percent).

All the broad emission lines exhibit very similar profiles and
 have nearly identical line
widths of 4000 to 4800 \kms{} (FWHM) except Ly$\alpha, $ which shows
a slightly broader line
width of 6300\,\kms. This difference might be caused by a varying continuum
or because of an 
additional underlying component in the Ly$\alpha$ line. 
The width (FWHM) of our recent H$\beta$ line measurement
 is narrower by 500 \kms with respect
to the spectrum taken by Boroson \& Green \cite{boroson92} in 1990. However, our
value corresponds with that of Ho \& Kim \cite{ho09} in 2004.
The differences in line width might have been caused by line variations in
the meantime.

%profile and additional underlying components in Ly$\alpha$ line
%i.e. $\ion{Si}{iii}\,\lambda 1206$ (see Fig.~\ref{pg0043_hst_2013_1991.ps})
%We show in
%Fig.~\ref{pg0043_velo_profile_cyclo2_cont2.ps} the profiles
%of the strongest
%UV lines that we attribute to the cyclotron lines CycA6, CycB4,
%and CycA5 -- in velocity space -- in comparison to the
%\ion{O}{vi}\,$\lambda 1038$ line.
%Again they are normalized to the same maximum intensity.
 All the 
cyclotron lines show very similar line widths of about 10\,000 \kms,
on the one hand, and they have entirely different line shapes
in comparison to those of the normal emission lines (FWHM$\sim5000~$\kms),
on the other hand.

The [\ion{O}{ii}]$\lambda$3727 line is the only
narrow forbidden  emission line detected in PG\,0043+039.
This [\ion{O}{ii}]$\lambda$3727 line exhibits a line width (FWHM) of 6.6\,\AA,{}
corresponding to a velocity of 530  \kms (see Fig.~\ref{pg0043_CaII.ps}). 
The Balmer lines exhibit the same systemic redshift as this narrow
emission line $[\ion{O}{ii}]\,\lambda 3727$.
The \ion{O}{vi}\,$\lambda 1038$  shows  an internal redshift
of 390 $\pm{}$ 150 \kms. However, this might be feigned because of 
absorption of the blue \ion{O}{vi} doublet component
(see Fig.~\ref{pg0043_velo3x2.ps}).
The Ly$\alpha$ line is definitely blueshifted
by 860 $\pm{}$ 150 \kms.

\subsection{UV/Opt. absorption line systems in PG\,0043+039}

PG\,0043+039 has been classified 
 as a BAL quasar (Turnshek et al.\citealt{turnshek94})
based on a broad CIV absorption at a blueshift of $\sim$ 10\,000 \kms.
We searched in our far-UV spectrum for additional absorption lines
belonging to commonly expected absorption line systems of, for example, 
 the $\ion{O}{vi}\,\lambda 1038$,
% $\ion{P}{v}\,\lambda 1128$, 
$\ion{N}{v}\,\lambda 1243$, $\ion{Si}{v}\,\lambda 1403$,
 and Ly$\alpha$ lines etc.
The $\ion{C}{iv}\,\lambda 1550$ line profile seen in our spectrum taken in 2013
is very noisy
and only covers  the blue wing 
at the outermost edge of our HST spectrum. The $\ion{C}{iv}\,\lambda 1550$ line profile,
 taken with the HST
in 1991, is overlayed in Fig.~\ref{pg0043_velo3x2.ps}.
We show in Fig.~\ref{pg0043_velo3x2.ps}
possible absorption profiles belonging to the OVI, NV, SiIV,
 CIV, MgII, and Ly$\alpha$ lines.
The profiles are divided  by the assumed continuum.
% The intensities of their emission profiles are normalized to 1 at
% $v_{\text{shift}} = 0 $ \kms. 
The velocity scale is calculated relative to the reddest
line of each multiplet.
The stronger blue component at about -1\,000 \kms is missing in the
 $\ion{O}{vi}$\,doublet ($\lambda 1032,1038$). This points to an absorption
 component at v $\sim$ 0 \kms. The same behavior is indicated in the other
high-ionization lines of the $\ion{C}{iv}\,\lambda 1550$ and
$\ion{N}{v}\,\lambda 1243$ doublets.
Additional possible positions of
absorption lines belonging to prominent
UV lines are indicated with vertical tick marks at
-20\,000, -16\,000, -13\,700, -11\,000, -9\,000 \kms.
Here we labeled local minima in the spectrum shortward
 of the emission lines.
However, no absorption troughs could be unambiguously connected to
any of the emission lines except for the CIV absorption.
   \begin{figure}
    \includegraphics[width=8cm,angle=0]{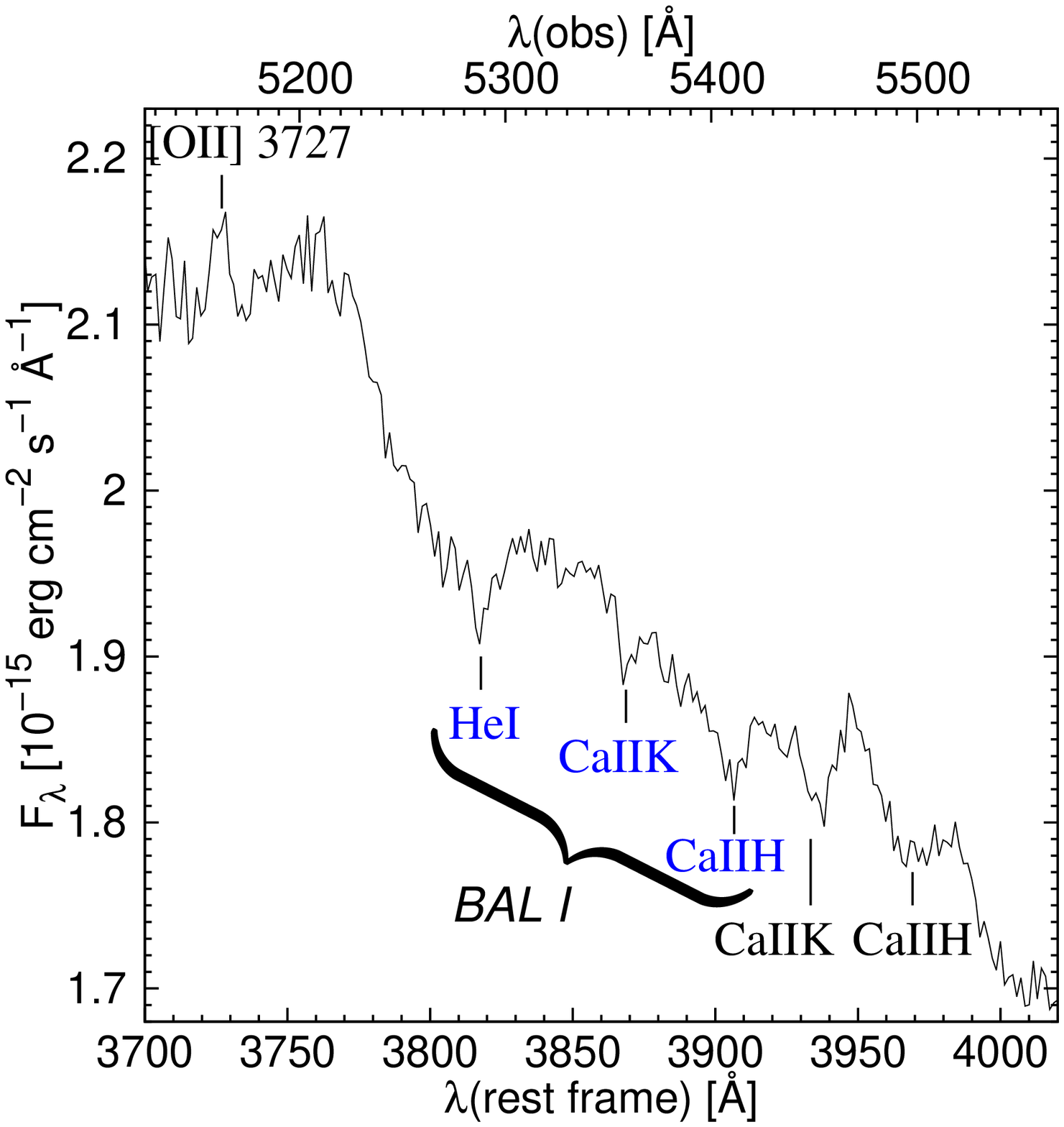}
      \caption{Enlargement of the optical rest-frame spectrum 
      around the narrow [\ion{O}{ii}]$\lambda$3727 emission and
     CaII H and K absorption lines. The BAL~I system (blue) consists of
 the CaH\,$\lambda 3968$, CaK\,$\lambda 3934$ lines
  (blueshifted by $\sim$ 4900 \kms) and the
\ion{He}{i}\,$\lambda 3889$ line (blueshifted by $\sim$ 5600 \kms).
              }
       \vspace*{-3mm} 
         \label{pg0043_CaII.ps}
%   \end{figure}
%
%
%   \begin{figure}
    \includegraphics[width=6.5cm,angle=270]{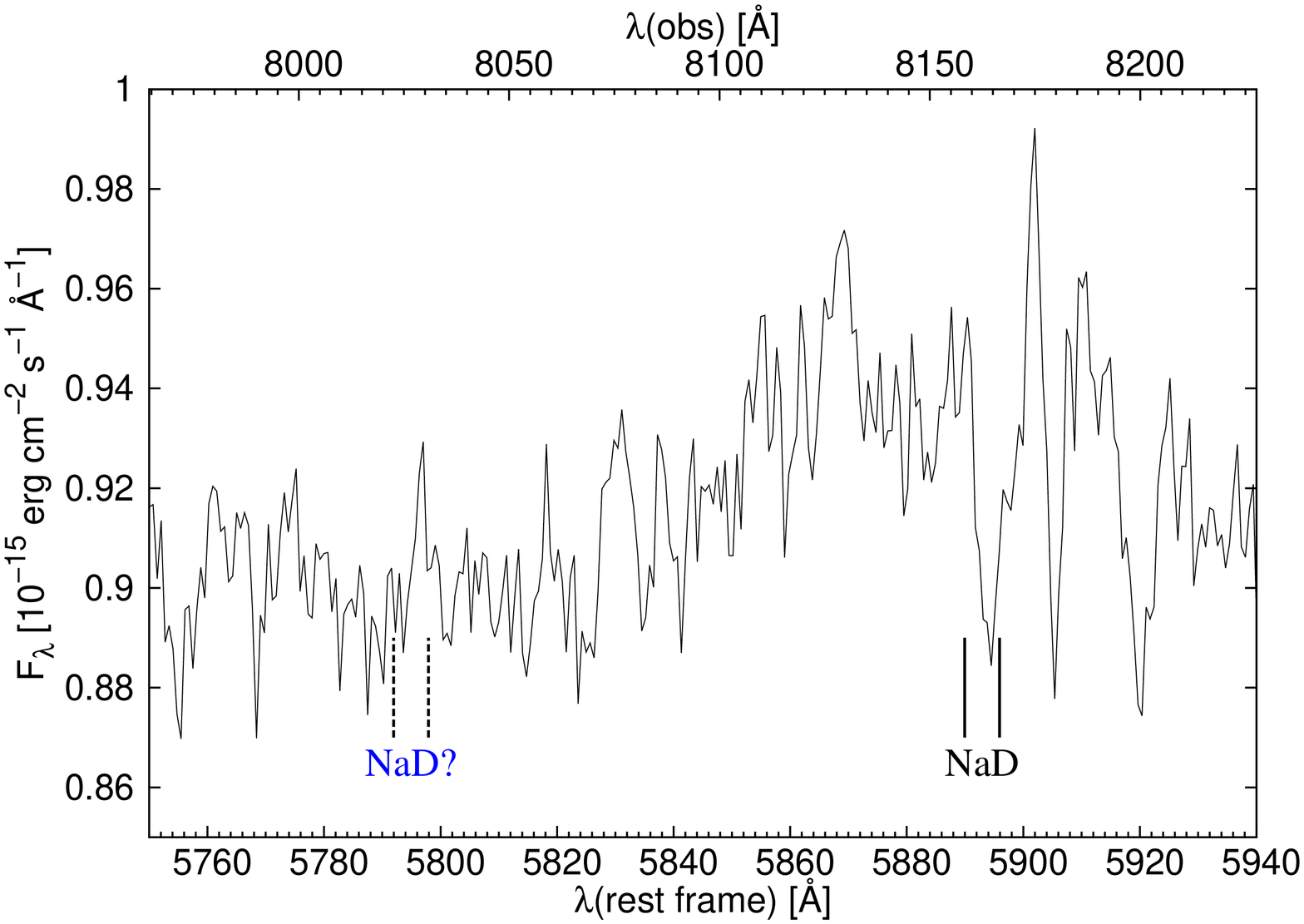}
      \caption{Enlargement of the optical rest-frame spectrum 
      around the intrinsic NaD absorption lines as well as around
      a possible blueshifted NaD absorption.
              }
       \vspace*{-3mm} 
         \label{pg0043_NaD.ps}
   \end{figure}
\begin{figure}
\centering
    \includegraphics[width=7.5cm,angle=0]{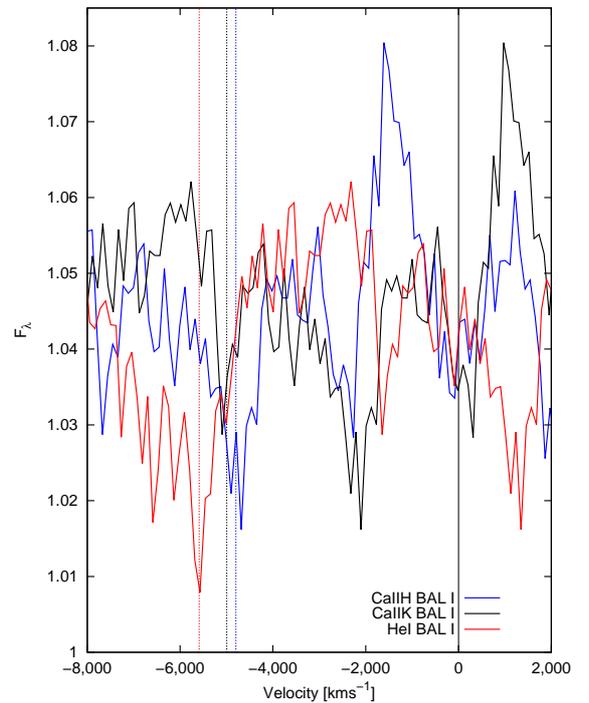}
      \caption{Absorption lines of the BAL~I system (CaH\,$\lambda
        3968$, CaK\,$\lambda 3934$, and \ion{He}{i}\,$\lambda 3889$ lines) in
        velocity space.} 
         \label{pg0043_velo_bal1.ps}
\end{figure}
A further aspect of our present study is the detection
of additional blueshifted BALs in the optical spectrum
of PG\,0043+039.
In 
Fig.~\ref{pg0043_CaII.ps} we present an extract of our optical spectrum
in the range of the narrow [\ion{O}{ii}]$\lambda$3727 emission line as well as
 CaII H and K absorption lines.
Figure~\ref{pg0043_NaD.ps}
 shows an enlarged section of our optical spectrum of PG\,0043+039
      around the NaD absorption region.
We are able to identify
weak absorption lines of Na\,D\,$\lambda 5890/96$, and the
CaH\,$\lambda 3968$ and CaK\,$\lambda 3934$ lines at the
systemic velocity of the galaxy. Their absorption line equivalent widths
are listed in Table~3.
\begin{table*}
%\centering
%\tabcolsep+3.5mm
\tabcolsep+5.5mm
\caption{Optical/UV absorption line equivalent widths and their shifts.}
\begin{tabular}{llllcc}
\hline 
\noalign{\smallskip}
Absorption Lines              &$W_{\lambda}$  & \multicolumn{2}{c}{Pseudo-continuum}      & Shift &comments\\ 
                              &[\AA{}]      &  blue side [\AA{}] &red side [\AA{}]     & [\kms]      &\\
\noalign{\smallskip}                                                                      
(1)                           & (2)         &                (3) &         (4)          &(5)    &(6)\\
\noalign{\smallskip}
\hline 
\noalign{\smallskip}
Na\,D\,$\lambda 5890/96$     &  .34$\pm{}$  .04  &5888 &5900  & 0   &\\
CaH\,$\lambda 3968$          &  .097$\pm{}$ .01  &3962 &3977  & 0   &\\ 
CaK\,$\lambda 3934$          &  .19 $\pm{}$ .05  &3929 &3944  & 0   &\\ 
\noalign{\smallskip}                                   
CaH\,$\lambda 3968$          & .11$\pm{}$ .03    &3899 &3912  & -4800 $\pm{}$ 300 & BAL\,1\\ 
CaK\,$\lambda 3934$          & .10$\pm{}$ .03    &3865 &3877  & -5000 $\pm{}$ 200 & BAL\,1\\ 
\ion{He}{i}\,$\lambda 3889$  & .25$\pm{}$ .05    &3808 &3828  & -5600 $\pm{}$ 200 & BAL\,1\\ 
\noalign{\smallskip}                                   
MgII\,$\lambda 2798$:        &  34.0$\pm{}$6.    &2531 &2779  & -19000 $\pm{}$ 1000 &       \\
CIV\,$\lambda 1550$ (broad)  &  19.2$\pm{}$1.    &1456 &1537  & -11100 $\pm{}$  1000 &       \\
CIV\,$\lambda 1550$ (narrow) &  .48$\pm{}$.1     &1542 &1552  & -800 $\pm{}$  400 &       \\
\noalign{\smallskip}
\hline
\end{tabular}\\
\end{table*}

  \begin{figure*}
\centering
    \includegraphics[width=10cm,angle=270]{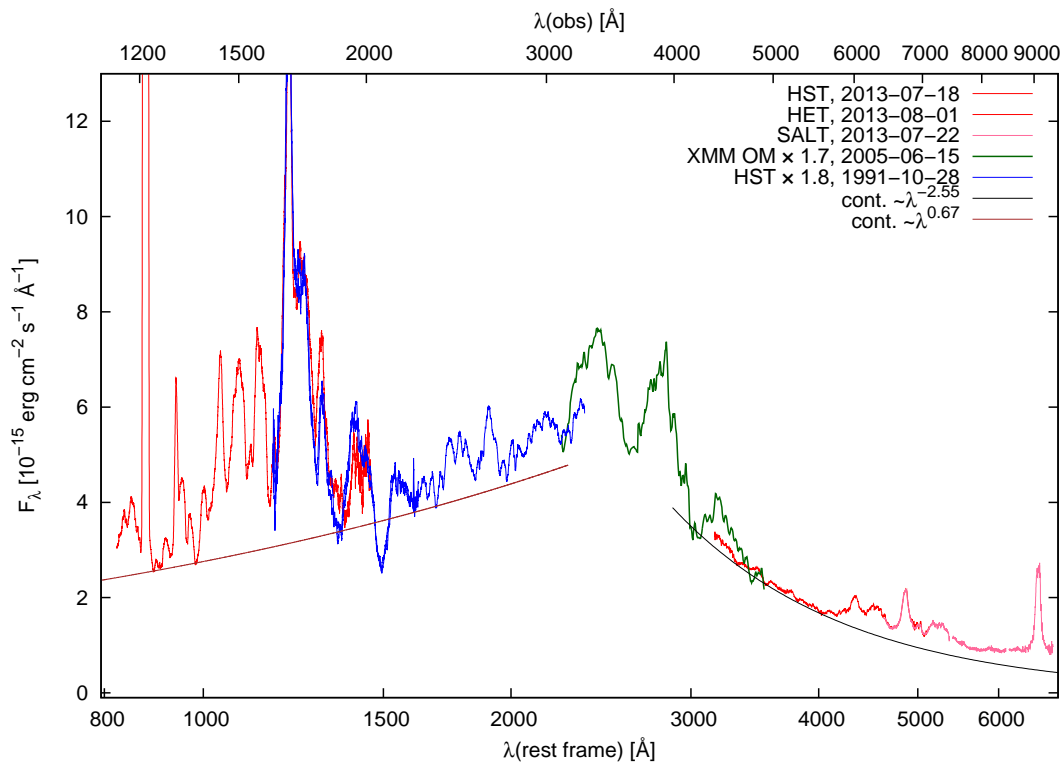}
      \caption{Combined optical-UV spectra of PG\,0043+039, corrected
for Galactic reddening, taken
in the years 2013 (red lines) and 1990/1991 (blue lines).
In addition the spectral OM data are given 
 taken with the XMM-Newton satellite in 2005 (green
data). Furthermore, the power-law continua to the UV and optical spectral ranges
are shown.}
%       \vspace*{-3mm} 
         \label{pg0043_sed_dered_log.ps}
   \end{figure*}

In addition to the absorption system at the systemic velocity,
 we can verify an absorption system (BAL~I)
consisting of the CaH\,$\lambda 3968$, CaH\,$\lambda 3934$,
 and \ion{He}{i}\,$\lambda 3889$ lines
(see Fig.~\ref{pg0043_CaII.ps}).
Figure~\ref{pg0043_velo_bal1.ps} shows the lines of the BAL~I system
in velocity space. The CaH\,$\lambda 3968$ and  CaH\,$\lambda 3934$ lines
are  blueshifted by 4900 \kms and the
\ion{He}{i}\,$\lambda 3889$ line is blueshifted by
$\sim$ 5600 \kms.
A similar absorption line system at a related
blueshift has been found before
in the peculiar BAL galaxy Mrk\,231 (Boksenberg et al.\citealt{boksenberg77}).
In this galaxy 
 even three BAL systems have been identified at different velocities
 and they vary independently
(e.g., Kollatschny et al.\citealt{kollatschny92}, 
Lipari et al.\citealt{lipari09}).
However, in contrast to Mrk\,231 we see
 no indication for a Na\,D
absorption at a blueshift of $\sim$ 5100 km $s^{-1}$ within the error limits.

\subsection{Overall spectral slope in PG\,0043+039}

We present    
a combination of our optical and UV spectra of PG\,0043+039 taken
in the summer of 2013 in Fig.~\ref{pg0043_sed_dered_log.ps}.
In addition a transition spectrum is overplotted taken
with the coaligned 30-cm optical/UV telescope (OM) on board 
 the XMM-Newton satellite in 2005.
We multiplied the observed OM spectrum with a factor of 1.7 to align
it with the optical and UV spectra taken in 2013.
In addition, we present the power-law continua adapted
 to the UV as well as optical spectral ranges.
There is a clear maximum in the overall continuum flux 
at around 2500 \AA{} (rest frame).
The flux is getting weaker toward shorter and longer wavelengths.
We compute a very steep UV power-law slope parameter
$\alpha_{ox}$=$-$2.55 $\pm{}$ 0.3 based on the near- and  far-UV data.

A maximum in the transition region between the optical and UV spectral
range has been noted before by
Turnshek et al.\cite{turnshek94}. 
They assigned this slope to intrinsic
reddening  of E(B-V) = 0.11 mag from dust similar to that found in the Small
Magellanic Cloud (SMC). 

   \begin{figure}
    \includegraphics[width=9.0cm,angle=0]{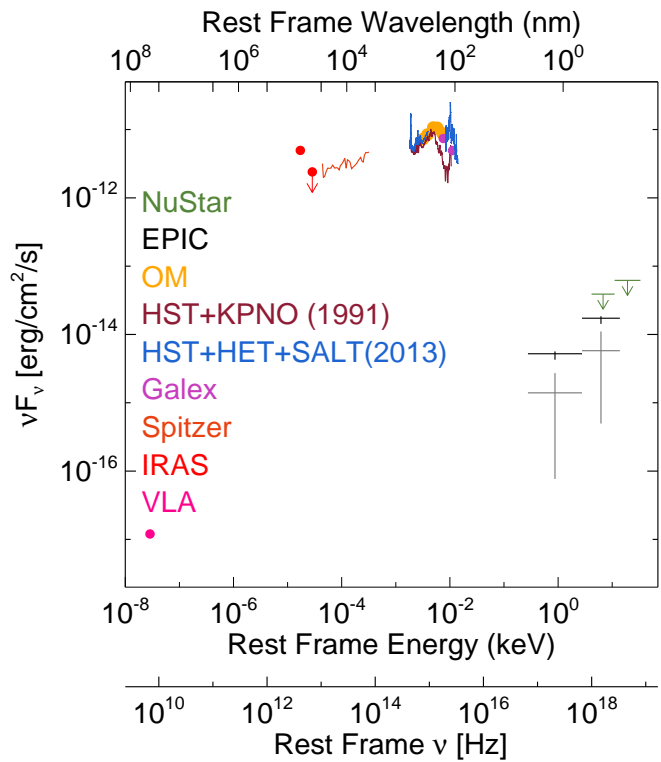}
      \caption{Combined overall spectrum (radio to hard X-ray) for PG\,0043+039
              }
       \vspace*{-3mm} 
         \label{pg0043_sed_15_3_3.ps}
   \end{figure}

The multiwavelength spectral energy distribution of PG\,0043+039 
from the 6cm radio (VLA) to the hard X-ray (NuStar) frequencies
is shown in Fig.~\ref{pg0043_sed_15_3_3.ps}. The 
radio observations with the VLA taken in  1982
 were published by Kellermann et al.\cite{kellermann89}.
In addition to its faintness in X-ray wavelengths,  PG\,0043+039 is a weak radio source.
This behavior is consistent with the general radio/X-ray luminosity
relation for radio-quiet quasars (Laor \& Behar\citealt{laor08}). 
Serjeant \& Hatziminaoglou\cite{serjeant09}
presented the infrared IRAS observations at 60 and 100 $\mu$ taken in 1983.
The Spitzer infrared and GALEX UV data were taken from their archives.

\section{Discussion}

\subsection{Continuum and line intensity variations}

We observed variations in the optical, UV, and X-ray continuum flux
of PG\,0043+039.  
Our observations show an increase in the optical and
UV continuum fluxes by a factor of 1.8 for  2013
compared to observations taken in 1990/1991.
Turnshek et al.\cite{turnshek94} reported optical and UV decreasing
flux variations of PG\,0043+039 at earlier epochs
with respect to their spectra taken in 1991. This means that PG\,0043+039
was in a low state in 1991.
PG\,0043+039 was by a factor of $\sim$1.6 brighter in the optical in
 1981 in comparison to their observations in 1990.
The same is true for the UV continuum flux in 1986.  

PG\,0043+039 has varied in the X-ray continuum as well.
While this object was not  verified in the X-ray in  2005 
in the first analysis  (Czerny et al.\citealt{czerny08}),
it now has been proven in 2013.
The flux increased
by a factor of 3.8~$\pm$0.9.
The X-ray variations of PG\,0043+039 were on the same order or even stronger
in comparison with long-term X-ray variations
of other BAL quasars
(see Saez et al.\citealt{saez12}). 
For example, Mrk\,231 varied in the X-ray by a factor 1.3 $\pm{}$ 0.2 over
a period of 20 years. 

Details of the spectroscopic variations in PG\,0043+039 seem to be even more
complex in comparison to their general continuum variations.
In the UV difference spectrum, based on HST spectra taken in 1991 and 2013
(see Fig. 6), individual emission lines stick out
after correction for a general intensity that increases by a factor of 1.8.
These lines have not been seen in the 1991 UV spectrum.
In addition, there are evident variations in
some broader UV structures (in intensity and wavelength).
 This variability behavior is different with respect
to what is known from normal spectral variations of AGN ( e.g.,
Kollatschny et al.\citealt{kollatschny14}).
A detailed analysis of these variations in PG\,0043+039 is beyond the scope of this paper.

\subsection{Emission lines in PG\,0043+039}

The Balmer lines H$\alpha$, H$\beta$, and H$\gamma$
(see Fig.~\ref{pg0043_het_salt.ps}) are the strongest emission lines
in the optical
spectrum of PG\,0043+039 besides strong FeII blends.
Again the optical spectrum of PG\,0043+039 shows similarities
to that of Mrk\,231, however, the 
relative strength of the optical FeII blends is even stronger in  Mrk\,231 
(Boksenberg et al.\citealt{boksenberg77};
Lipari et al.\citealt{lipari09}) when compared to PG\,0043+039.
The line intensity ratio of the broad H$\alpha$ to H$\beta$ lines 
 has a value 2.38
(Table~1) and cannot be understood with simple photoionization models
in which values of 2.8 or more are expected.  

The [\ion{O}{ii}]$\lambda$3727 line is the only verified
narrow forbidden  emission line in PG\,0043+039.
We can verify this line in our optical spectrum in contrast to 
Turnshek et al.\cite{turnshek94} who reported their absence.
%Ho \& Kim (\citealt{ho09}) took an optical spectrum of PG\,0043+039 with the
%Magellan 6.5m Telescope in 2004. They reported an uncertain intensity value
%for $[\ion{O}{ii}]\,\lambda 3727$.
% Was brighter in the continuum (at 5100:-14.09) and H$\beta$ (-12.31). 
The [\ion{O}{ii}]$\lambda$3727 line is the only detected
forbidden line in Mrk\,231 as well
(Boksenberg et al.\citealt{boksenberg77};
Lipari et al.\citealt{lipari09}).
Based on the fact that there are only upper limits
for the [\ion{O}{iii}]$\lambda$5007 line intensities in both galaxies,
we find that
the intensity ratio  [\ion{O}{ii}]$\lambda$3727/[\ion{O}{iii}]$\lambda$5007
is higher in PG\,0043+039 and Mrk\,231 in comparison to normal AGN.
It is known that there is an anticorrelation between the strengths
of the [\ion{O}{iii}]$\lambda$5007 line and the
FeII blends (Eigenvector 1; e.g., Sulentic et al.\citealt{sulentic00}).
However, an analogical anticorrelation of the 
[\ion{O}{ii}]$\lambda$3727/[\ion{O}{iii}]$\lambda$5007 ratio
 has not been studied yet.

Typically, the
Ly$\alpha$, \ion{O}{vi}\,$\lambda 1038$, and \ion{N}{v}\,$\lambda 1243$ lines
%(see  Fig.~\ref{pg0043_hst.ps}).
are the strongest emission lines in the UV range
from 800 to 1500\,\AA{}  in AGN.
We derived line intensity ratios for these emission lines in PG\,0043+039
(see Table 1 and Fig.~\ref{pg0043_shull_hst_cos.ps}),
%the \ion{O}{vi}\,$\lambda 1038$/Ly$\alpha$/\ion{N}{v}\,$\lambda 1243$   
which are very similar to those seen in mean composite AGN spectra
(Shull et al.\citealt{shull12}). However, PG\,0043+039 shows additional
strong broad lines in its FUV spectrum, which 
 could not be attributed to known emission lines.

\begin{figure*}
\centering
    \includegraphics[width=9.8cm,angle=270]{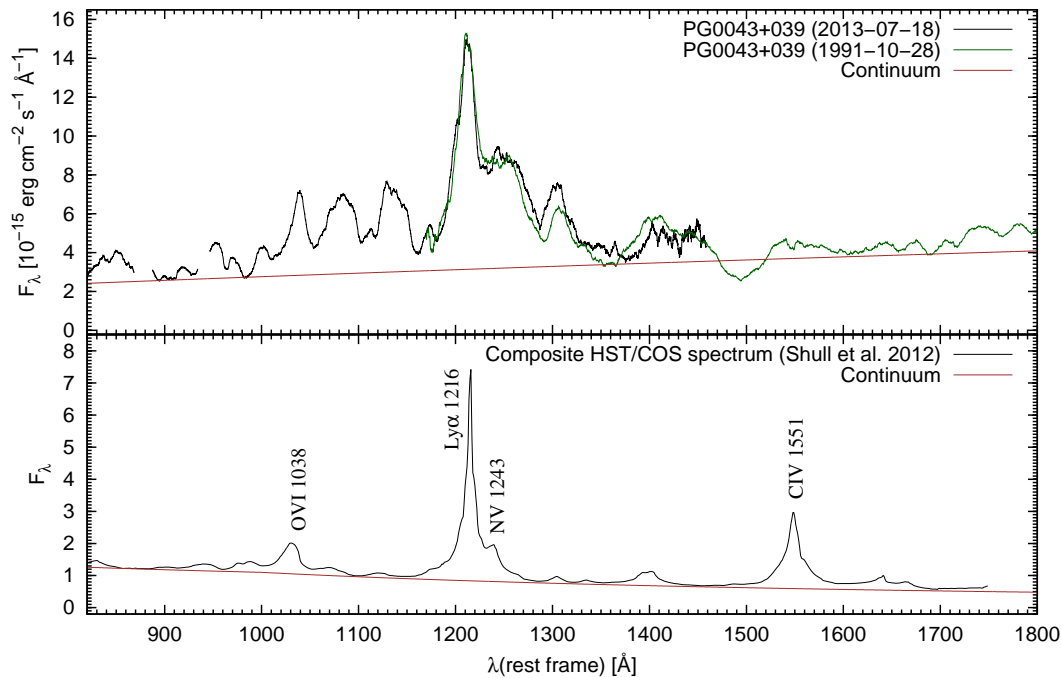}
      \caption{Comparison of the  UV emission line intensities in 
PG\,0043+039 (upper panel) with respect to a composite HST/COS spectrum of 22 AGN
(Shull et al.\citealt{shull12}, lower panel). We combined the two
ultraviolet spectra of PG\,0043+039 taken with the HST in
the years 2013 and 1991. The UV line and continuum intensities increased
by a factor of 1.8 between these two observations. We multiplied the UV
spectral flux taken in 1991 with this factor to match the UV observations taken
in 2013. The continua of both spectra are indicated as
brown lines. }
       \vspace*{-3mm} 
         \label{pg0043_shull_hst_cos2x1.ps}
   \end{figure*}

\begin{figure*}
\centering
    \includegraphics[width=9.8cm,angle=270]{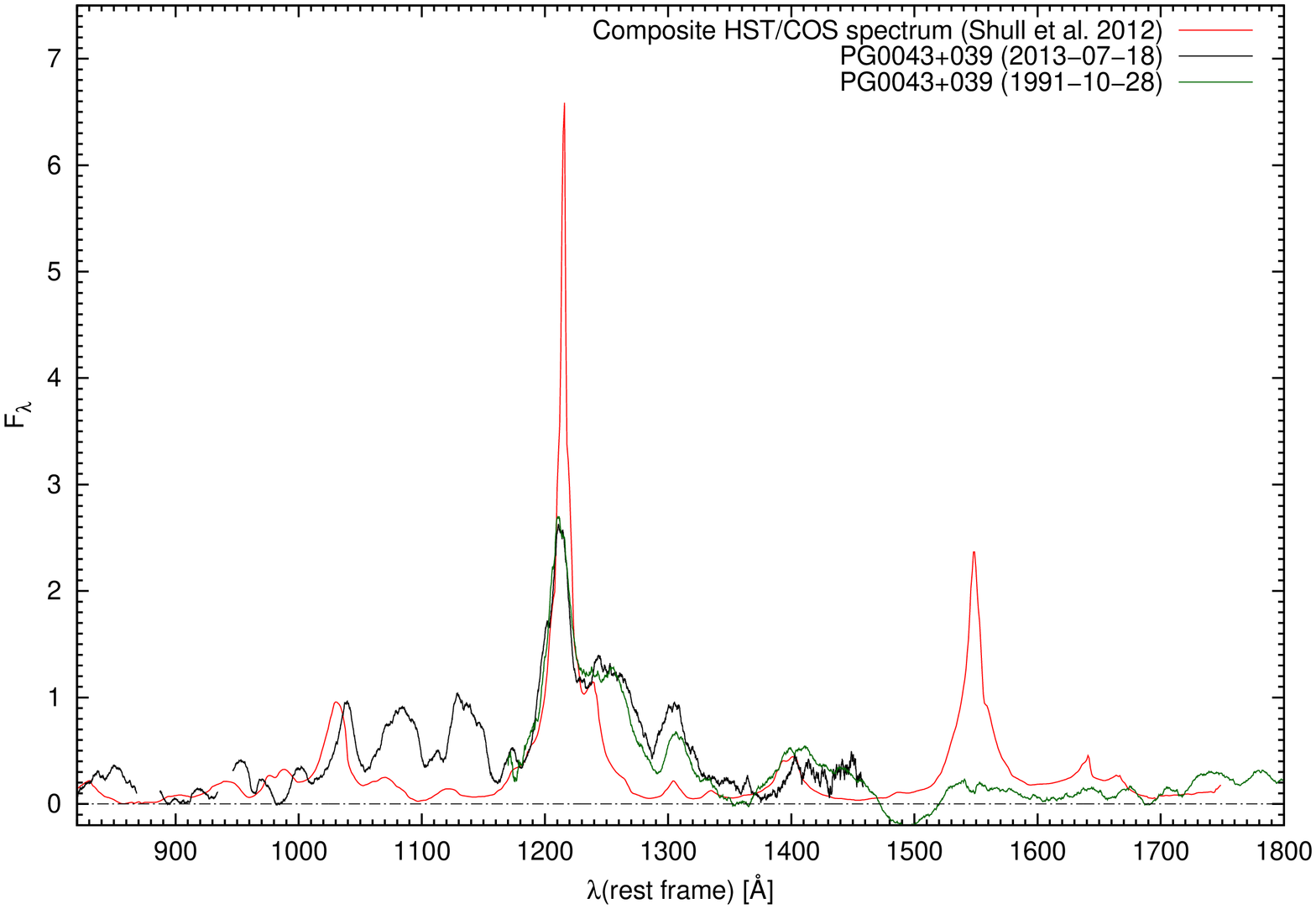}
      \caption{Comparison of the relative UV emission line intensities in 
PG\,0043+039 with respect to a composite HST/COS spectrum of 22 AGN
(Shull et al.\citealt{shull12}). The continua have been subtracted first and
both spectra are scaled to the same $\ion{O}{vi}\,\lambda 1038$
and $\ion{N}{v}\,\lambda 1243$ line intensities.
              }
       \vspace*{-3mm} 
         \label{pg0043_shull_hst_cos.ps}
   \end{figure*}

\begin{figure*}
\centering
    \includegraphics[width=10cm,angle=270]{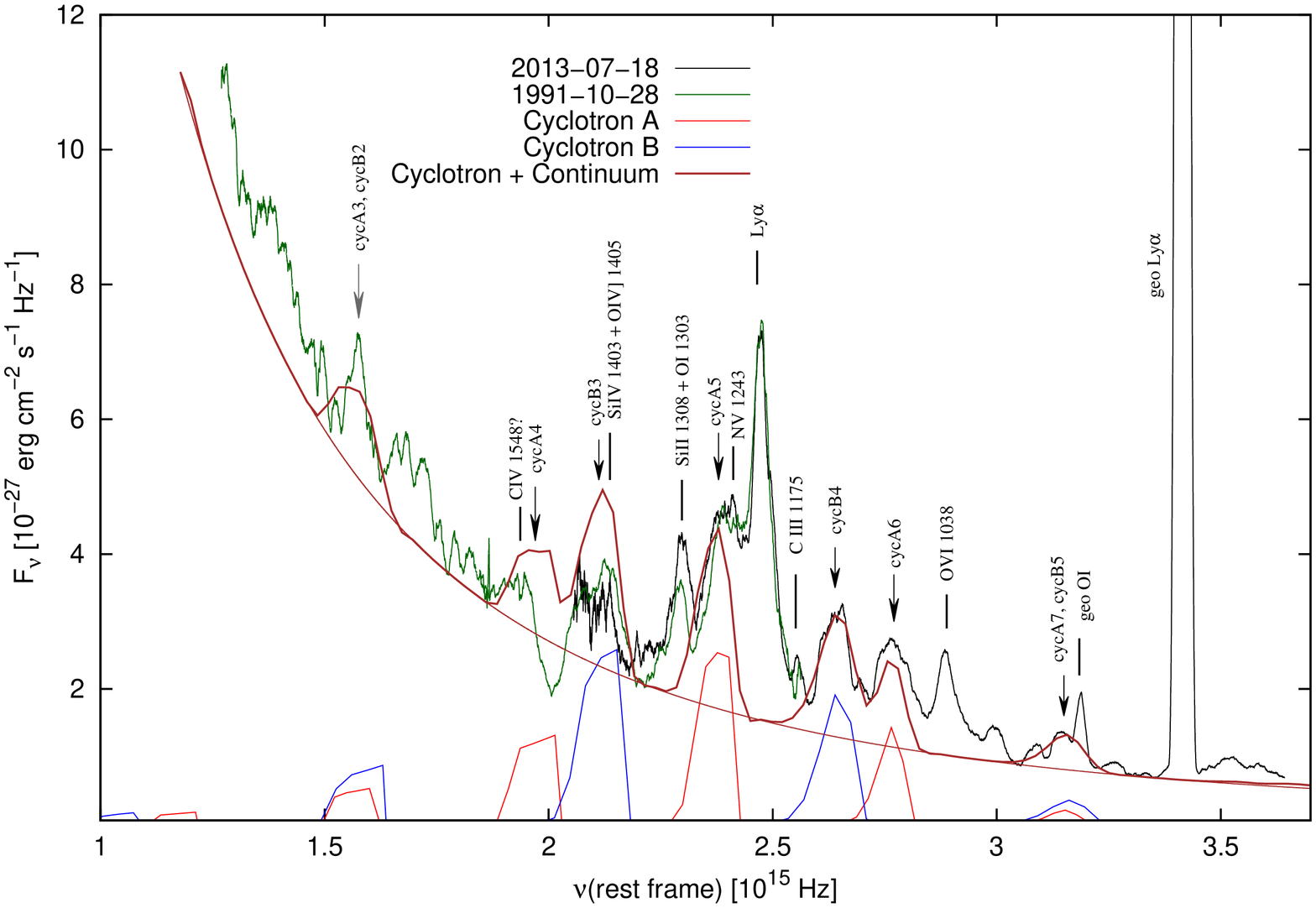}
      \caption{Combined ultraviolet spectra of PG\,0043+039 taken with the HST 2013 and 1991. We indicae  the identifications of the strongest
UV emission lines,  the geo-coronal lines, and  both cyclotron
systems B and A with
their 3rd to 7th harmonics. The integer numbers identify the
emission humps with multiple of the cyclotron fundamental.
We indicate the power-law continuum N$_\nu\sim\nu^{\alpha}$ with
$\alpha = 0.69 \pm 0.02.$
The modeling of both cyclotron line systems B and A is given at the bottom.
              }
       \vspace*{-3mm} 
         \label{pg0043_cyc_nu.ps}
   \end{figure*}

We present in Fig.~\ref{pg0043_shull_hst_cos2x1.ps} the combined UV-spectrum
of PG\,0043+039 and a composite HST/COS spectrum of 22 non-BAL quasars
taken from Shull et al.\citealt{shull12}. The UV continua of both spectra
show different gradients
and emission line spectra.
We compare in Fig.~\ref{pg0043_shull_hst_cos.ps} the relative UV emission
line intensities in 
PG\,0043+039 with respect to the composite HST/COS AGN spectrum.
We subtracted the continua  first and
scaled both spectra  to match the $\ion{O}{vi}\,\lambda 1038$
and $\ion{N}{v}\,\lambda 1243$ line intensities by means of
 a single scaling factor.  
Both spectra
are similar with respect to the relative intensities of the
strongest emission lines, $\ion{O}{vi}\,\lambda 1038$, Ly$\alpha$ 
and $\ion{N}{v}\,\lambda 1243,$ as well as their equivalent widths
(see Table~4). However, they differ with respect to
the \ion{C}{iv}\,$\lambda 1550$ line absorption
and with respect to the strong broad humps.
In Paper I we presented  a modeling of these observed strong humps in
PG\,0043+039 by means of cyclotron lines.
They are not seen in the composite AGN spectrum.
We derived plasma
temperatures of T $\sim$ 3~keV
and magnetic field strengths of  B $\sim$ 2 $\times10^{8}$ G
for the cyclotron line-emitting regions close to the black hole.
 With this modeling, we could explain the
wavelength positions of the broad humps and their relative intensities.
Figure~\ref{pg0043_cyc_nu.ps} shows the UV spectrum of PG\,0043+039
in frequency space with the identifications of the cyclotron systems A and B 
with their second to seventh harmonics in addition to the normal emission lines
(see Paper I for more details). 

Independent indications for magnetic fields of at least tens of Gauss (and
possibly considerably higher) on scales on the order of light days from
a central black hole have been recently reported by Marti-Vidal
\cite{marti-vidal15}. They detected Faraday rotation at sub-mm wavelengths
in the nearby AGN 3C84.

PG\,0043+039 is a very luminous AGN  (M$_B$=$-26.11$) compared to Mrk\,231
 (M$_B$=$-21.3$).
%On the other hand the relative intensities of the Balmer lines with
%respect to Ly$\alpha$ are more normal in PG\,0043+039.
However, the blueshift of the  Ly$\alpha$ line is stronger
in Mrk\,231 ($-$3500 \kms; Veilleux et al. \citealt{veilleux13})
 than in  PG\,0043+039 ($-$860 \kms).
Typical AGN show blueshifts of $-$400 \kms.
%
%Furthermore there is no indication for
%$\ion{N}{v}\,\lambda 1243$ emission in Mrk\,231.
Some additional information is contained in the line profiles as well.
Normally, the broad AGN emission lines show different line widths
(Mrk110; e.g., Kollatschny\citealt{kollatschny03a,kollatschny03b})
caused by their formation at different distances
from the central ionizing source.
Besides that, as an example
 the AGN Mrk\,231 shows a peculiar Ly$\alpha$ profile
compared to the Balmer lines (Veilleux et al.\citealt{veilleux13}).
In contrast to that, all the broad emission lines in PG\,0043+039 show similar
line profiles and similar line widths
(see Fig.~2 in Paper I)
independent of their ionization state.
This is an indication that the broad emission line region
in PG\,0043+039 is not strictly
 structured and that
 all these lines originate at similar distances
from the central ionizing source.
%\ion{O}{vi}\,$\lambda 1038$ ionization potential : 114 eV in comparison to 
%\ion{He}{ii}\,$\lambda 4686$ ionization potential : 54.4 eV
%Very high ionization potential for \ion{O}{vi} .

\subsection{Absorption line systems in PG\,0043+039}

 Bahcall et al.\cite{bahcall93}
and Turnshek et al.\cite{turnshek94,turnshek97} classified PG\,0043+039
 as a BAL quasar
based on a broad CIV absorption line at a blueshift of $\sim$ 10,000 \kms
detected with the HST in 1991.
This study did not find convincing evidence for BALs due to low-ionization transitions of
AlII or CII, other than for MgII. 
On the other hand, only about 10 percent of the BAL quasars show 
these lines (e.g., Trump et al.\citealt{trump06};
Allen et al.\citealt{allen11}).
Bechtold et al.\cite{bechtold02} reanalyzed
the absorption line spectra of all quasars
taken with the high-resolution gratings of the FOS 
on board of the HST. They could not confirm the existence
of broad absorption lines in PG\,0043+039 as reported before by 
 Turnshek et al.\cite{turnshek94}. They only verified narrow
absorption in the Ly$\alpha$ and \ion{C}{iv}\,$\lambda 1550$ lines at the same
redshift as the emission lines.  
Based on the UV absorption lines profiles in Fig.~\ref{pg0043_velo3x2.ps},
as well as on the UV spectral distribution in Fig.~\ref{pg0043_sed_dered_log.ps},
we confirm a central absorption component at v $\sim$ $-$800 \kms 
in the high-ionization doublet lines $\ion{O}{vi}$, $\ion{C}{iv}$,
$\ion{N}{v}$,
and, in addition, a strong, broad
absorption component at $-$11\,000  \kms  in the \ion{C}{iv}\,$\lambda
1550$ line (see Table 3).
The possible identification of a strong, broad absorption component belonging
to the
MgII\,$\lambda 2798$ line at $-$19\,000  \kms  cannot
be unambiguously demonstrated.
There is a strong absorption structure in the the UV/optical
spectral distribution at 2620~\AA{} rest frame
(see Fig.~\ref{pg0043_sed_dered_log.ps}).
However, this structure might not be a real absorption line and it  might 
only be simulated by broad emission line humps
at shorter and longer wavelengths.
%on both sides (see Fig.~\ref{pg0043_sed_dered_log.ps}).

The single clear identification of the broad
 \ion{C}{iv}\,$\lambda 1550$ absorption at $-$11\,000  \kms
in PG\,0043+039 is atypical compared to the UV absorption
properties in other BAL 
quasars (see, e.g., Baskin et al.\citealt{baskin13};
Hamann et al.\citealt{hamann13}).
In general, BAL 
quasars also show absorption features
from other high- and low-ionization lines,
such as OVI and Ly$\alpha$.
On the other hand, the UV spectrum of PG\,0043+039
shows similarities with the UV spectrum of Mrk\,213.  
Veilleux et al.\cite{veilleux13} demonstrated the surprising
absence of UV absorption in the FUV spectrum in Mrk\,231.
 Ly$\alpha$ and \ion{C}{iv}\,$\lambda 1550$ absorption has only been
unambiguously identified
 in Mrk\,231. In PG\,0043+039, 
 not even an indication of Ly$\alpha$
 absorption has been seen.

%discussion about absorption systems: in comparison to
%Here we discuss whether some of the absorption
%gaps between the bumps are connected with absorption profiles.
%Mrk\,231 is an AGN showing unknown broad bumps in the UV as well (Veilleux
%et al. \citealt{veilleux13})

In the optical spectral range, we verified a BAL~I system at a blueshift
 of $\sim$ 4900 \kms in the CaH\,$\lambda 3968$ and CaK\,$\lambda 3934$
lines and at a blueshift of $\sim$ 5600 \kms in the
\ion{He}{i}\,$\lambda 3889$ line.
A similar system with similar shifts has been identified before
in the peculiar BAL galaxy Mrk\,231 (Boksenberg et al.\citealt{boksenberg77}).
The velocity of the high-ionization line \ion{He}{i} is slightly
higher than that of the low-ionization \ion{Ca}{ii} lines. It has been proposed
by Voit et al.\cite{voit93} that the low-ionization lines could be produced
in dense cores with relatively lower velocity, while the high-ionization lines
are produced in a thinner wind, possibly ablated from the cores and accelerated
(see also the discussion in Leighly et al.\citealt{leighly14}).
%However, the equivalent widths of the absorption line systems
%are far more stronger
%in Mrk\,231 than in  PG\,0043+039.
In Mrk\,231
even three BAL systems at different velocities have been found
 and these systems varied independently
(e.g., Kollatschny et al.\citealt{kollatschny92}; 
Lipari et al.\citealt{lipari09}).

In PG\,0043+039 there is no indication for a Na\,D
BAL absorption at a blueshift of $\sim$ 5100 km $s^{-1}$ 
within the error limits. This
nondetection of Na\,D BAL absorption is an important clue since
Na\,D BAL absorption requires
dust (e.g., Veilleux et al.\citealt{veilleux13}). This strengthens the model
that dust absorption is not responsible for the weak FUV flux in PG\,0043+039.

\subsection{Opt/UV/X-ray slope}

\begin{figure}
    \includegraphics[width=10.cm,angle=0]{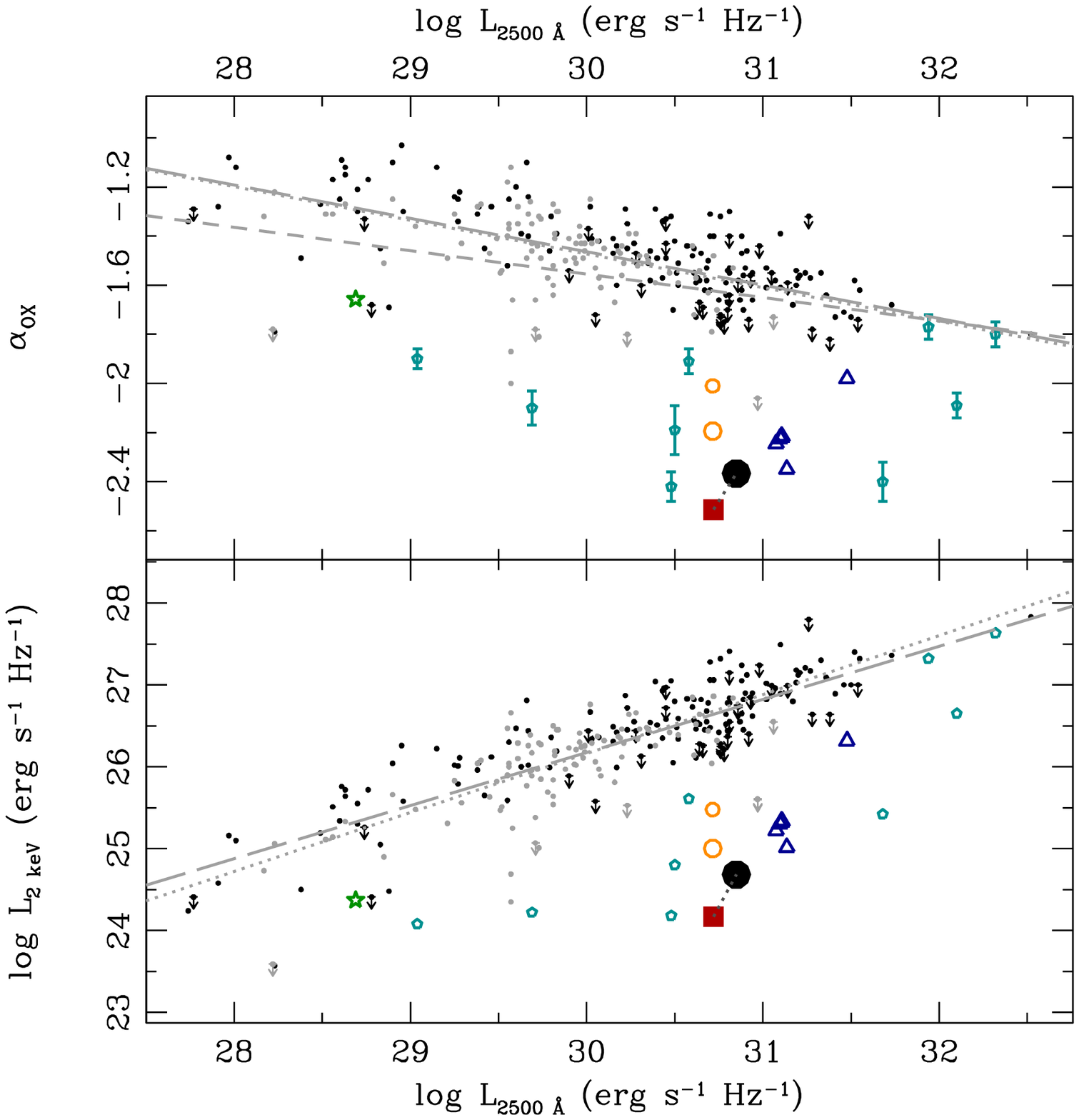}
      \caption{The $\alpha_{ox}$ gradient as a function of the monochromatic
luminosity at rest-frame $2500\,$\AA{} (upper panel) and the monochromatic
luminosity at 2 keV versus the monochromatic
luminosity at $2500\,$\AA{} (lower panel). 
Black-filled circles indicate the SDSS objects with 0.1$\leq$z$\leq$4.5 from
 Strateva et al. \cite{strateva05}, while gray-filled circles label the
data from Steffen et al.\cite{steffen06}.
The dotted line represents the best fit to the sample of Steffen et
al.\cite{steffen06}. 
Further measurements of extreme X-ray weak quasars
are indicated with  cyan circles (Saez et al.\citealt{saez12}).
 The positions of PG\,0043+039 are highlighted for
the years 2005 (red square)
and 2013 (black circle).
              }
       \vspace*{-3mm} 
         \label{luvVSstra_weak_v2.ps}
   \end{figure}

Based on our simultaneous observations in different frequency ranges,
we could show that the X-ray faintness of PG\,0043+039 is intrinsic
and not simulated by variations.  
The $\alpha_{ox}$ gradient is shown in
Fig.~\ref{luvVSstra_weak_v2.ps} for a sample of AGN
as a function of the monochromatic
luminosity at rest-frame $2500\,$\AA{} (upper panel) and the monochromatic
luminosity at 2 keV versus the monochromatic
luminosity at $2500\,$\AA{} (lower panel). 
Black-filled circles indicate the SDSS objects with 0.1$\leq$z$\leq$4.5 from
 Strateva et al. \cite{strateva05}, while gray-filled circles label the
data from Steffen et al.\cite{steffen06}.
The long-dashed lines are the best-fit linear relations for the samples
by Strateva et al. \cite{strateva05}.
The dotted line represents the best fit to the sample of Steffen
et al.\cite{steffen06}. 
Further measurements of extreme X-ray weak quasars
(Saez et al.\citealt{saez12}) are incorporated by cyan circles.
Additional observations are given for the quasars PG~2112+059
(blue open triangle; Schartel et al. \citealt{schartel10},
\citealt{Schartel2007}),
PG~1535+547 (green star; Ballo et al. \citealt{ballo08}),
and PG~1700+518 (yellow open circle; Ballo et al. \citealt{ballo11}).
Among all galaxies PG\,0043+039 shows the most extreme $\alpha_{ox}$ gradient
($\alpha_{ox}$=$-$2.55) based on the data taken in  2005.
The X-ray flux of this AGN increased by a factor of 3.8 in 2013.
However, it still shows a very extreme
$\alpha_{ox}$ gradient ($\alpha_{ox}$=$-$2.37) at that epoch.
We presented the extreme UV/X-ray weakness of PG\,0043+039 compared to a mean
spectral distribution of QSOs (Richards et al.\citealt{richards06})  in Fig.~3 of Paper I.
% The mean spectral distribution of a QSO is very similar to the
%spectral energy distribution of type 1 AGN (Scott \& Stewart
%\citealt{scott14}).
%The overall spectrum of PG\,0043+039 is very similar to the spectrum of the
%X-ray weak quasar PG1700+518  (siehe in Luo, 2013) and CIV absorption. 

There is the question as to what causes the extreme X-ray faintness of
PG\,0043+039. We observed PG\,0043+039 simultaneously in the optical, UV, and
X-ray bands to exclude that an apparent X-ray weakness is
simulated by variations. Another possible explanation of  the extreme X-ray faintness
could be the absorption of the X-ray flux due to gas.
 However,
the X-ray faintness of PG\,0043+039 is consistent with the extrapolation
of its faint UV-flux (Fig.~3 of Paper I). Furthermore, the upper
limit seen in the hard X-ray flux, and confirmed by the simultaneous
 NuSTAR observations (Luo et al.\citealt{luo14}),
is a further indication of an intrinsically X-ray faintness 
 (Fig.~\ref{pg0043_sed_15_3_3.ps}).
Other arguments against X-ray absorption
come from the results of X-ray modeling;
the X-ray spectra show no sign of extreme absorption.
Intrinsically absorbed power-law fits give a power-law index
of 1.7.  This is compatible with the standard value 1.9
and N$_H$~$=$ 5.5~$\times$ 10$^{21}$~cm$^{-2}$,
which is not enough to explain
the extreme weakness of the quasar. Moreover, the explanation that the X-ray spectrum has a completely absorbed primary
continuum and an absorbed reflection  failed (Sect. 3.2).

A possible explanation for the
weakness of the X-ray flux might be the
suppression of a hot inner accretion disk in PG\,0043+039. 
It is generally accepted that the main contribution to the X-ray flux of AGN
is coming from a hot corona surrounding the inner accretion disk via
comptonization of UV photons from the disk
(Haardt \& Maraschi\citealt{haardt91}, Luo et al.\citealt{luo14}).
If there is no inner accretion disk, then
% because of suppression by strong magnetic
%fields as known from intermediate polar stars or AMHer stars (Lit.)
a strong X-ray emitting corona could not evolve. 
The link between the disk and the corona has also been
independently demonstrated 
by Gliozzi et al.\cite{gliozzi13}, based on an
AGN long-term monitoring campaign with Swift.

There are further indications for the nonexistence
of a hot inner accretion disk in
PG\,0043+039 (besides the X-ray weakness): the UV/FUV flux is suppressed in
this galaxy. There is a maximum in the UV continuum flux at around 
$\lambda  \approx 2500$\AA{}.
%(corresponding to a temperature of X K)
The flux decreases toward shorter wavelengths
in contrast to most other AGN in which a maximum is found at around
$\lambda  \approx 1000$\AA{} (e.g., Vanden Berk et
al.\citealt{vandenberk01}).
 A turnover at this wavelength corresponds to a 
maximal accretion disk temperature of $T_{\text{max}} \approx 50\,000$\,K
(Laor \& Davis\citealt{laor14} and
references therein). Therefore, a turnover at 2500 \AA{}
as seen in PG\,0043+039 should only correspond
to a temperature of about 20\,000 K.
 It is generally accepted that the observed
FUV continuum emission in AGN is produced by an accretion disk.

The weak UV flux in PG\,0043+039
has been interpreted by Turnshek et al.\cite{turnshek94}
as intrinsic reddening of SMC-like dust. However, their spectral fit
did not correspond with the observations
 with this kind of a reddening correction. Furthermore,
that assumption leads to a surprising low BAL region column density.
In the end, they admitted that intrinsic dust extinction might  not be the only
plausible explanation for the turndown in the continuum flux shortward
of 2200\AA{}.
Similarly, Veilleux et al.\citealt{veilleux13} tried to model the
turndown in the UV
continuum flux in Mrk\,231. However, another plausible explanation for
 the weak FUV continuum
flux might be the hypothesis that there is simply no FUV flux emitted because
of a nonexisting hot inner accretion disk.
In the same spirit,
 some quasars have been identified that show unusual weak blue continua
(e.g., Hall et. al.\citealt{hall02}; Meusinger et al.\citealt{meusinger12}).
We see no Lyman edge ($\lambda  < 912$\AA{} )
in our spectrum of PG\,0043+039, in accordance to the ultraviolet composite
spectrum of some AGN (Shull et al.\citealt{shull12}).
 Baskin et al.\cite{baskin13} also
saw no detectable Lyman
edge associated with the BAL absorbing gas.
%In general AGN show a break at ($\lambda  \sim 1000~\AA{} $)
%-- see the composite spectrum  of Shull et al.\cite{shull12}.
%However, the rest-frame continuum in PG\,0043+039 
%($N_{\nu} \sim  \nu^\alpha$) does not 
%show such a break at ($\lambda  \sim 1000~\AA{} $) in contrast
%to composite AGN spectra.

It is known from cataclysmic stars as AMHer stars, so-called polars,
as well as from intermediate polars that they host  strong magnetic
fields. Their magnetic fields, on
the order of $\times10^{8}$ G, are responsible for the prevention of the
formation of an (inner) accretion disk in these objects.
 Analogically,
the expansion of magnetic bridges between the ergosphere and the disk around 
rapidly rotating black holes could be responsible for an outward shift
of the inner accretion disk (Koide et al.\citealt{koide06}).

Another indication of an accretion disk that is not that extended  comes from the
widths of the Balmer, Ly$\alpha$, and 
\ion{O}{vi}\,$\lambda 1038$ lines (see Paper I).
Typically, the broad emission lines have different widths,
as they originate at different distances from the central ionizing
source (e.g., in Mrk\,110, Kollatschny\citealt{kollatschny03a,kollatschny03b}).
However, all the low- and high-ionization lines in PG\,0043+039 show
the same widths, indicating
that they originate at the same distances from the central ionizing region.

%A final indication comes from the broad unidentified humps in the 
%UV spectrum. These humps have broader profiles in comparison to
%the regular emission line profiles (see
%Fig.~\ref{pg0043_velo_profile_cyclo2_cont2.ps})
%and there is no much evidence that these humps are simulated
%by broad absorption lines (see Fig.).

\begin{table}
%\centering
\tabcolsep+5.8mm
%\tabcolsep+1.3mm
\caption{Comparison of the
  composite spectrum of "normal" AGN  (Shull et al.\citealt{shull12}) with the
  spectrum of PG\,0043+039. Equivalent widths of selected UV lines.}
\begin{tabular}{lrr}
\hline 
\noalign{\smallskip}
Emission Lines                &\multicolumn{2}{c}{Equivalent width [\AA{}]}\\ 
                              &Shull+12 & PG\,0043+039 \\
\noalign{\smallskip}
(1)                           & (2)   & (3)\\
\noalign{\smallskip}
\hline 
\noalign{\smallskip}
\ion{O}{vi}\,$\lambda 1038$  & 23  $\pm{}$ 3 & 26  $\pm{}$ 3\\
Ly$\alpha$                   & 115 $\pm{}$ 5 & 120 $\pm{}$ 5 \\
\ion{N}{v}\,$\lambda 1243$   & 23  $\pm{}$ 2 & 34  $\pm{}$ 3 \\
\noalign{\smallskip}
\hline
\end{tabular}
\end{table}

\section{Summary}

We took deep X-ray spectra  with the XMM-Newton satellite,
 FUV spectra with the HST,
 and optical spectra of PG\,0043+039 with the 
   HET and SALT telescopes in July, 2013.
PG\,0043+039 is one of the weakest  quasars in the X-ray. We barely
detected this object in our new deep X-ray exposure.
It has
an extreme  $\alpha_{ox}$  continuum gradient of $\alpha_{ox}$=$-$2.37.
However, the X-ray spectra show no sign of extreme absorption.
Moreover, an attempt to
  explain the X-ray spectrum with a completely absorbed primary
  continuum and an absorbed reflection has failed.

PG\,0043+039 shows a maximum in the overall continuum flux at around 
$\lambda  \approx 2500$\AA{}  
in contrast to most other AGN where a maximum is found at 
shorter wavelengths. In combination
with its intrinsic X-ray weakness
this is an indication for an accretion disk that is not that hot
compared to most other AGN.

PG\,0043+039 has been classified as a BAL
 quasar before, based on a broad
CIV absorption. We detected no further absorption lines in our
FUV spectra. However, in the optical we found a narrow  
BAL system in the CaH\,$\lambda 3968$, CaK\,$\lambda 3934$ lines
(blueshifted by 4900 kms$^{-1}$), and in the  
\ion{He}{i}\,$\lambda 3889$ line (blueshifted by 5600 kms$^{-1}$).

The UV/optical flux of PG\,0043+039 has increased by a factor of 1.8
compared to spectra taken in  1990/1991. Some UV emission lines
appeared in the new UV spectrum taken in 2013, which were not
present in the spectrum from 1991.
In addition, the UV spectrum is highly peculiar showing no 
Lyman edge. Furthermore, strong broad humps are to be seen
 in the FUV, which  have not been
identified before in other AGN.
We modeled the observed strong humps in the
UV spectrum of PG\,0043+039
by means of cyclotron lines.
We derived plasma
temperatures of T $\sim$ 3~keV
and magnetic field strengths of  B $\sim$ 2 $\times10^{8}$ G
for the cyclotron line-emitting regions close to the black hole (see Paper I).

\begin{acknowledgements}
This work has been supported by the DFG grant Ko 857/32-2.

Some of the observations reported in this paper were obtained with
the Southern African Large Telescope (SALT).
The Hobby-Eberly Telescope (HET) is a joint project of the University of
 Texas at Austin, the Pennsylvania State University, Stanford University,
 Ludwig-Maximilians-Universit\"at M\"unchen, and Georg-August-Universit\"at
G\"ottingen.

This research has made use of the NASA/ IPAC Infrared Science Archive,
which is operated by the Jet Propulsion Laboratory, California Institute
of Technology, under contract with the National Aeronautics and Space
Administration.
Some GALEX data presented in this paper were obtained from the
Mikulski Archive for Space Telescopes (MAST). STScI is operated by the
Association of Universities for Research in Astronomy, Inc., under NASA
contract NAS5-26555. Support for MAST for non-HST data is provided by
the NASA Office of Space Science via grant NNX09AF08G and by other
grants and contracts.

\end{acknowledgements}

\end{document}